\documentclass[aps,prb,twocolumn,showpacs]{revtex4-1}

\usepackage{bm}
\usepackage{graphicx}
\usepackage{amsmath}
\usepackage{amssymb}

\begin{document}
\title{Bulk-boundary correspondence in topological systems with the momentum dependent energy shift}
\author{Huan-Yu Wang}\email{eurake27hywang@fzu.edu.cn}
\affiliation{Fujian Key Laboratory of Quantum Information and Quantum Optics and Department of Physics,
\\Fuzhou University, Fuzhou 350108, China}
\author{Zhen-Biao Yang}
\affiliation{Fujian Key Laboratory of Quantum Information and Quantum Optics and Department of Physics,\\
Fuzhou University, Fuzhou 350108, China}
\author{Wu-Ming Liu}
\affiliation{Beijing National Laboratory for Condensed Matter Physics, Institute of Physics, Chinese Academy of Sciences, Beijing 100190, China}
\begin{abstract}
Bulk-boundary correspondences (BBCs) remain the central topic in modern condensed matter physics, and are gaining increasing interests with the recent discovery of non-Hermitian skin effects. However, there still exist profound features of BBCs that are beyond the existing framework. Here, we report the unexpected behavior of BBC when the Hamiltonian contains term of the form $d_0(k) I$, which serves as a momentum dependent energy shift. For Hermitian cases, momentum dependent energy shift can force the system to be semimetal, where topological phase transitions can take place with the upper and the lower bands keeping untouched.  The proper modified BBC should be reconstructed from the perspective of the direct band gap. In non-Hermitian cases,  skin effects are found to be capable of coexisting with the preserved BBC, of which the process can be greatly facilitated by the complex $d_0(k)I$. Remarkably, such results can be led a further step and contrary to the intuitive consideration, the modified BBC in Hermitian systems can be restored to be conventional by including extra non-Hermiticity. The physical origin for these phenomena lies in that  $d_0(k)I$ can drastically change the point gap topology. Finally, the corresponding experimental simulations are proposed via the platforms of electric circuits.
\end{abstract}
\maketitle
\section{Introduction}
Topological nontrivial materials featuring isolated in-gap edge modes are gaining wide applications in various platforms, such as cold atom systems \cite{PhysRevLett.105.255302,advances,Natphys,PhysRevLett.111.120402,PhysRevA.89.022319,PhysRevLett.103.020401,PhysRevB.89.174501} and optical waveguide arrays \cite{Natpho,PhysRevB.102.161112,PhysRevLett.104.150403,Nature,PhysRevA.104.043513,PhysRevB.102.100303,PhysRevLett.111.103901,PhysRevLett.126.113902,PhysRevLett.111.103901,PhysRevB.105.195129,PhysRevLett.124.083603}. One of the most typical ways to characterize topological nontrivial states of matter is achieved via the bulk topological invariants obtained with periodical boundary conditions. Conventionally, preserved BBC suggests that non-vanishing  bulk topological invariants indicate the presence of nontrivial edge modes in open boundary conditions \cite{PhysRevB.100.174512,PhysRevB.104.165125,PhysRevB.105.064304,Science}.

However, with the recent development of non-Hermitian physics, it has been found that not only the edge modes but  all the bulk states can pile up at the edges of the systems as well and such phenomenon is dubbed as non-Hermitian skin effects \cite{PhysRevLett.123.246801,PhysRevLett.129.113601,Natphys2,PhysRevLett.124.066602,PhysRevLett.121.086803,PhysRevLett.125.186802,PhysRevLett.125.126402,PhysRevA.107.062214,Natcommu, Natcommu2,PhysRevLett.124.086801,PhysRevB.102.205118,PhysRevResearch.5.013185,PhysRevA.107.043315}. In these cases, the bulk boundary modes can be extremely sensitive to local perturbations and the conventional BBC can be broken \cite{PhysRevLett.127.116801,PhysRevA.95.023836}. Meanwhile, non-Hermitian topological edge modes may not remain degenerate but collapse to a single one, becoming exceptional for Hamiltonian of not full rank \cite{RevModPhys.93.015005,PhysRevB.101.045130,PhysRevLett.123.066405,PhysRevB.107.075118,PhysRevLett.123.066405,PhysRevLett.126.230402,PhysRevA.98.023818,PhysRevB.103.125302}. Consequently, the original topological invariants may fail to describe the non-Hermitian nontrivial edge modes. To retain the  proper topological characterization, the normal Bloch wave vector has to be deformed to be complex and the generalized Brillouin zone is named accordingly \cite{PhysRevLett.123.066404,JPCM2,PhysRevResearch.1.023013}.

Despite of the achieved results so far, it has been discovered that BBC still share profound features beyond the existing framework. In this work, we fill in this blank by demonstrating how BBC behaves when the Hamiltonian contains terms of the form $d_0(k)I$ in both Hermitian and non-Hermitian cases. Such structures can be deemed as the momentum dependent energy shifts which usually appear in systems considering the inter-cell tunneling or the site-dependent chemical potential.  First, for Hermitian cases, we present that topological phase transitions can take place with the upper and the lower bulk band keeping untouched. At this stage, the conventional gap closing points may not characterize the emerging of topological edge modes properly. Indeed,  the momentum dependent energy shift can force the system to be semimetal  and  topological features can only be captured by the information of the direct band gap, which measures the gap amplitude between the minimum of upper band and the maximum of lower band.

For non-Hermitian systems, we exhibit that the complex $d_0(k)I$ term can facilitate the presence of discontinuous skin effects which coexist with the preserved BBC. Specifically, in this process, the determinant of transfer matrix remains identity and  bulk states are localized at different ends of the lattice. Here, it shall be pointed out that such skin effects differ from the $Z_2$ skin effects by the absence of Kramers pair considering the symmetry constraints. Meanwhile, we reveal the physical origin for the phenomenon above lying in that $d_0(k)I$ can drastically influence the point gap topology, which manipulate the shape of generalized Brillouin zone, and shift the energy spectrum at the same time.  Remarkably, these results can be led a further step to the nonintuitive phenomenon that the modified BBC in Hermitian systems with semimetals can be restored to be conventional  by introducing extra non-Hermiticity. As a detailed illustration, in one dimension, we consider the concrete example taking the form of modified Su-Scrieffer-Heeger (SSH) chain and demonstrate the bulk invariants signifying topological phase transitions.   Finally, the experimental simulations to show topological edge modes with preserved BBCs are proposed via the  platform of electric circuits \cite{Natcommu3,PhysRevA.106.052216,PhysRevB.100.045407,Natphys3,PhysRevLett.122.247702}.
\section{Effects of $d_0(k)I$ in 1D topological nontrivial lattices}
\subsection{ Modified BBC in Hermitian semimetals induced by $\mathbf {d_0(k) I}$}
First, we are to demonstrate how BBC shall be constructed when coming across purely real $d_0(k)I$ in Hermitian systems. As a concrete example, we consider the following modified SSH chain
\begin{eqnarray}
\begin{aligned}
H&=\sum^{N}_{i=1} t_1(c^{\dagger}_{i,A}c_{i,B}+c^{\dagger}_{i,B}c_{i,A})+\sum^{N-1}_{i=1}t_2 (c^{\dagger}_{i,B}c_{i+1,A}\\
&+c^{\dagger}_{i+1,A}c_{i,B})+iJ (c^{\dagger}_{i,A}c_{i+1,A}-c^{\dagger}_{i+1,A}c_{i,A})\\
&+iJ(c^{\dagger}_{i,B}c_{i+1,B}-c^{\dagger}_{i+1,B}c_{i,B}),
\end{aligned}
\end{eqnarray}
where $J,t_{2}\in \mathrm{R}$ denote the inter-cell tunneling between the same and different types of sublattices respectively. $t_1$ describes the intra-cell tunneling. $N$ is the length of the chain. The momentum space Hamiltonian can be obtained by applying  the spatial Fourier transformation $c_{k,A(B)}=\frac{1}{\sqrt{N}}\sum_i c_{i,A(B)} e^{-ikR_{i,A(B)}}$ and takes the form
\begin{figure*}[t]
\centering
\includegraphics[width=0.99\textwidth,height=0.39\textheight]{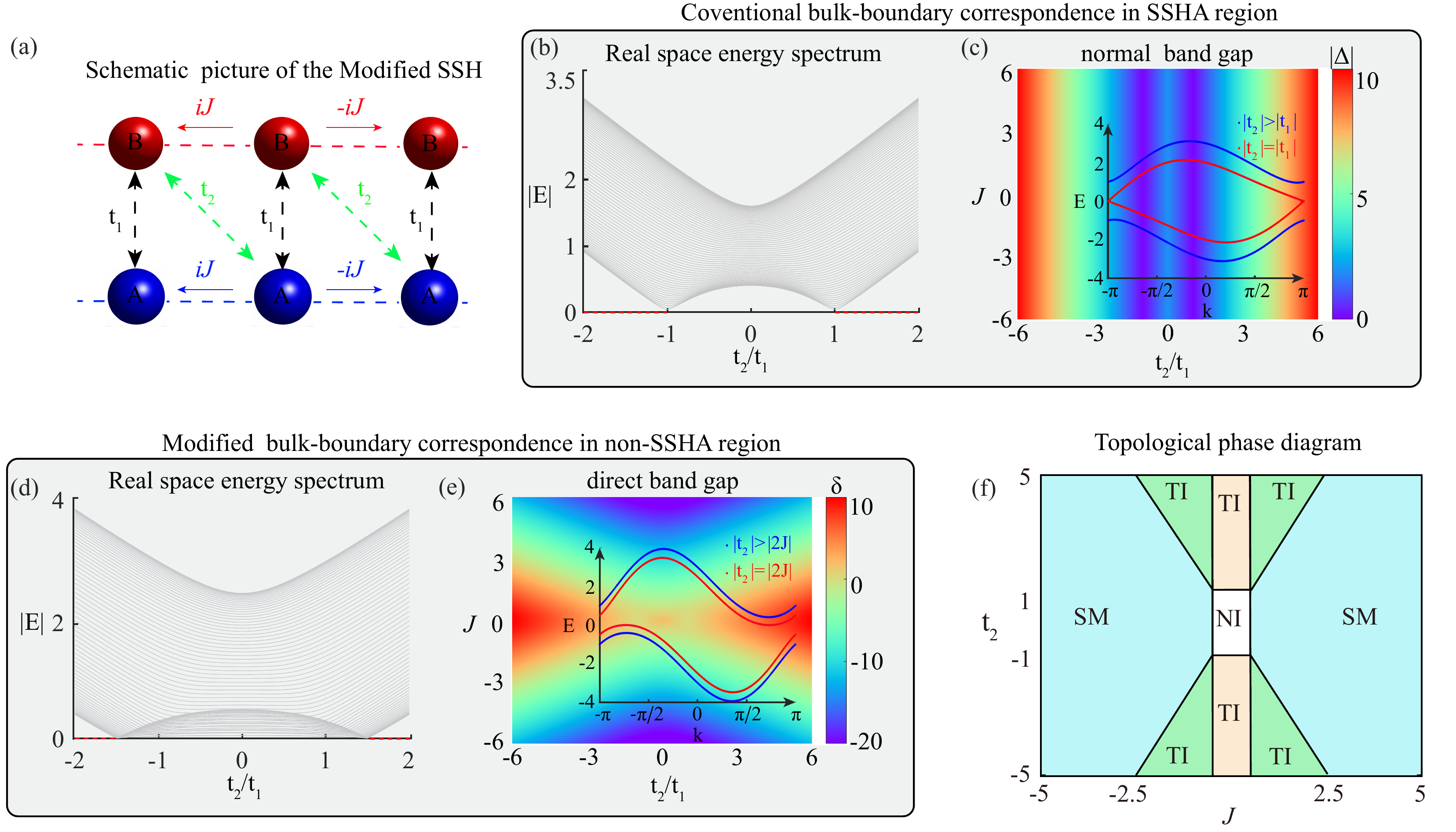}
\caption{(a) The schematic picture of the modified SSH chain. (b) When $|2J|<|t_1|$, the system appears to be  SSHA and topological edge modes marked with red dashed line lie at $|t_2|>|t_1|$. (c) The absolute value of band gap amplitude $|\Delta|$ varies as a function of $t_2/t_1$ and $J$. The subfigure depicts that corresponding to (b), the system is a gapped TI for $|t_2|>|t_1|$ and becomes gapless at $|t_2|=|t_1|$. The conventional bulk-boundary correspondence is fulfilled. (d) When $|2J|>|t_1|$, topological nontrivial region exists with $|t_2|>|2J|$.  (e) The direct band gap amplitude $\delta$ varies as a function of $t_2/t_1$ and $J$. The subfigure demonstrates that when the inter-cell tunneling is enlarged to $|2J|>|t_1|$, the system is a TI with $|t_2|>|2J|$ and turns to a semimetal (SM) at $|t_2|=|2J|$. At this stage, the modified bulk-boundary correspondence should be constructed from the perspective of the direct band gap. (f) Topological phase diagram as a function of inter-cell tunneling $J$ and  $t_2$. In contrast to  SSH model, not all regions satisfying $|t_2|>|t_1|$ are topological nontrivial. The orange area depicts the SSHA  TI and the green  area describes the non-SSHA TI.  } \label{fig1}
\end{figure*}
\begin{eqnarray}
\begin{aligned}
H(k)\!\!&=\!\!-2J\sin{(k)}I\!+\![t_1\!+\!t_2\cos{(k)}]\sigma_x\!+\!t_2\sin{(k)}\sigma_y\\
&=d_0(k)I+d_x(k)\sigma_x+d_y(k)\sigma_y,
\end{aligned}
\end{eqnarray}
where $I$ and $\sigma_{x,y,z}$ are the identity matrix and pauli matrices. The first term on the right hand side, $d_0(k) I$, can shift the energy spectrum momentum dependently and force the system to be semimetal.  Eq. (2) indicates the absence of the time-reversal symmetry.  Nevertheless, the particle-hole symmetry is preserved as
\begin{equation}
\Xi H^{*}(k) \Xi^{\dagger}=-H(-k), \Xi=\sigma_z.
\end{equation}
Hence, our one-dimensional quantum chain belongs to the topological class D and the reasonable topological invariants are the $Z_2$ indexes. Meanwhile, it shall be noticed that when $k=0,\pi$, the $d_0(k),d_y(k)$ terms are vanishing and only $d_x(k)$ term survives. In consequence, the $Z_2$ topological invariants are defined as
\begin{eqnarray}
\begin{aligned}
Q=sgn[d_x(k=0)d_x(k=\pi)],
\end{aligned}
\end{eqnarray}
where the topological nontrivial regions are supposed to be signified by $Q=-1$ and $|t_1|<|t_2|$.

One illustration of the BBC manifests that the critical value of parameters, where the real space topological edge modes merge to the bulk, should be consistent with the value that the momentum space Hamiltonian becomes gapless. For our system schematically shown in Fig. 1(a)  with $t_1=1.0$, when the inter-cell tunneling $J<0.5=t_1/2$, the system appears to be SSH alike (SSHA) and topological nontrivial edge modes lie in the regions  $|t_1|<|t_2|$ [see Fig. 1(b)]. At this stage, the conventional BBC is fulfilled [see Figs. 1(b-c)], where the gapped topological insulator (TI) phases are marked by $Q=-1$, and the topological trivial phase depicts the normal insulator (NI) with $Q=1$.

However, as the inter-cell tunneling $J$ is further enlarged, the system exhibits distinguishable  topological properties compared to the original SSH model. In detail, the topological nontrivial region appears to be located at  $|t_2|>|max\{t_1,2J\}|$ and there exists topological trivial regions satisfying $|t_2|>|t_1|$ for $|2J|>|t_1|$ [see Fig. 1d]. At this stage, bulk  invariant $Q$ shall fail to characterize topological features. Specifically, from the perspective of the momentum space energy spectrum, it can be observed that topological phase transition can still take place while the upper bands do not touch the lower bands. We tend to term such cases to be  non SSH alike (non-SSHA).

Numerically, the results in Fig. 1(e) suggest that the proper modified BBC should be reconstructed by including the definition of the direct band gap $\delta$, which measures the energy difference between the minimum of upper band [$E^{u}_{min}(k_1)$] and the maximum of the lower band [$E^{l}_{max}(k_2)$], $\delta= E^{u}_{min}(k_1)-E^{l}_{max}(k_2)$. Here, we need to point out that $\delta$ has taken the momentum dependence of energy shift into consideration and as long as  $k_1\neq k_2$, the direct band gap differs from the normal band gap $\Delta$. Meanwhile, $\delta<0$ signifies the semimetal phase rather than the normal insulator phase. Consequently, transitions between the semimetal and topological insulators can properly be described by the gap closing points of the direct band gap. The topological phase diagram as a function of the inter-cell tunneling $t_2$ and $J$ is shown in Fig. 1(f).

\subsection{Preserved BBC with the presence of  non-Hermitian skin effects}
The previous discussion demonstrates how the BBC behaves when the Hermitian systems exhibit semimetal phases induced by $d_0(k) I$.  For open quantum systems with asymmetric tunneling or on-site gain-loss, the Hamiltonian may not remain Hermitian and can  feature the skin effects. Conventionally, the presence of non-Hermitian skin effects can be identified in two ways: a) the generalized Brillouin zone does not coincide with the original Brillouin zone and the momentum space energy spectrum $\{\mathrm{Re}(E_k),\mathrm{Im}(E_k)\}$ encircles non-zero area with the momentum $k$ running over the whole Brillouin zone; b) the determinant of transfer matrix is not identify. In the following, we are to present how the BBC will be  affected by the  complex $d_0(k) I $ terms.

One way to induce non-Hermiticity in our modified SSH chain can be achieved via changing the inter-cell tunneling between the same type of sublattices to be purely imaginary $J\rightarrow -iJ$, and the system  can be redescribed as
\begin{equation}
H_{nh}(k)=\sum_k 2iJ\sin{k} I+[t_1+t_2\cos(k)]\sigma_x+t_2\sin(k)\sigma_y.
\end{equation}
Eq. (5) suggests that the imaginary part of energy spectrum is only decided by the first term on the right hand side. Hence, $\mathrm{Im} (E_k)$ should always be gapless as long as $J\neq 0$. Considering the real part of the eigen-energy,  $\mathrm{Re} (E_k)$ is exactly the same as that of the SSH chain and should be gapped as long as  $|t_1|\neq |t_2|$. Hence, if the BBC  is fulfilled, topological edge modes shown by the absolute energy spectrum $|E|$ will appear (or merge to the bulk) at $|t_2|=|t_1|$.

Numerically, the energy spectrum with open boundary conditions for the model in Eq. (5) is illustrated by Fig. 2(a). Such results approved the preserved BBC.  However, in Fig. 2(b), we notice that the real space energy spectrum $\{\mathrm{Re}(E), \mathrm{Im}(E)\}$ obtained with open boundary conditions (red dashed line) does not coincide with that obtained by periodic boundary conditions (blue dashed line) in the large size limit, fixing $t_2=-0.5, J=1$. Meanwhile, in Fig. 2(c), the momentum space energy spectrum $\{\mathrm{Re}[E(k)], \mathrm{Im} [E(k)]\}$ encircles non-zero area with $k$ running over the Brillouin zone $[-\frac{\pi}{a}, \frac{\pi}{a}]$, of which the results, according to ref [28], suggest the presence of non-Hermitian skin effects.
\begin{figure*}[t]
\centering
\includegraphics[width=0.99\textwidth,height=0.399\textheight]{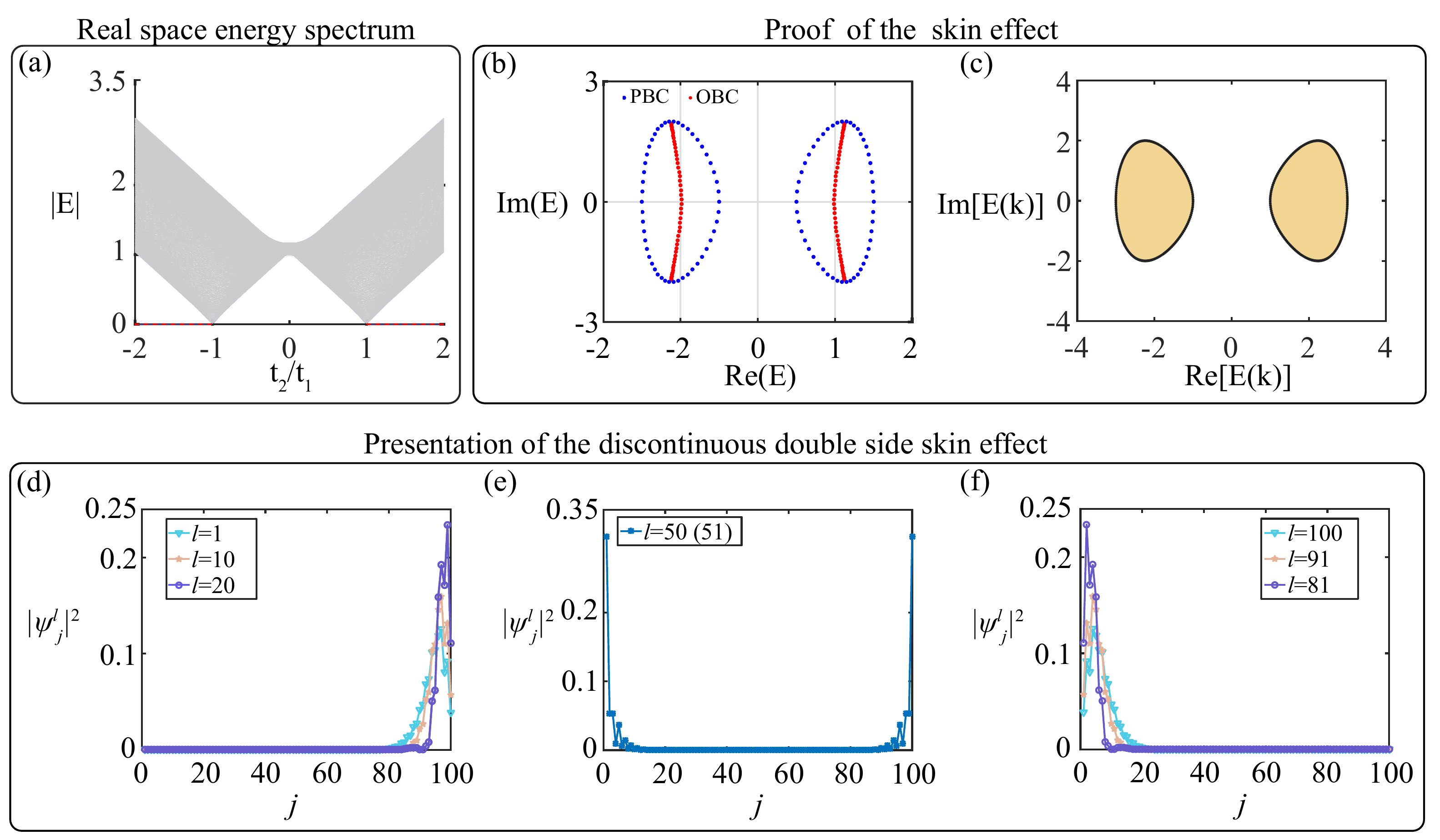}
\caption{(a) The energy spectrum of  modified NH-SSH chain in open boundary conditions varies as a function of $t_2/t_1$ with fixed $J=1$. Topological edge modes merge to the bulk at $|t_1|=|t_2|$.  (b) The real space energy spectrum shown by  $\{\mathrm{Re}(E), \mathrm{Im}(E)\}$ in open boundary conditions (red dashed line) does not coincide with that obtained with periodical boundary conditions (blue dashed line). (c) The momentum space energy spectrum $\{\mathrm{Re}[E(k)], \mathrm{Im}[E(k)]\}$ forms loops enclosing non-zero area when the translational momentum runs through the whole Brillouin zone. (d-f) The localization of the modified NH-SSH chain with $J=1, t_1=1, t_2=-2$. Considering the representative  bulk states, $|\psi^{l}_j|^2, l=1,10,20,100,91,81$ are not equally extended but localized at the right and left end of the lattice respectively, suggesting the double-side non-Hermitian skin effects. $\psi^{50(51)}_j$ is the combination of two zero energy edge states and has nearly equal localization on both ends, where no non-Hermitian skin effects shall exist.} \label{fig2}
\end{figure*}

Typically, non-Hermitian skin effects tend to result in the broken BBC and the discussions above are noticeably against this. To further identify the localization of all the eigenstates, we apply the transfer matrix approach, of which the method is entirely based on the real space Hamiltonian. Here, we assume the boundary modes to be $\psi=\sum_{p}\varphi^{N}_{p}, p=A,B,  \varphi^{N}_{P}=\beta^{N} \varphi^{0}_p$. The determinant of transfer matrix is equal to $|\beta|$ and describes how the states are boosted or suppressed during the wave propagating.  A demonstration on how the transfer matrix approach succeeds in identifying the skin effects in the non-Hermitian SSH chain is presented in the methods.  Specific to our modified model, we present that the determinant of transfer matrix is identity. In detail, given the basis $\Psi^{\dagger}_i=(c^{\dagger}_{i,A},c^{\dagger}_{i,B})$, the Hamiltonian can be depicted by
\begin{equation}
H_{nh}=\sum_i \Psi^{\dagger}_i A \Psi_i+\Psi^{\dagger}_i B \Psi_{i+1}+\Psi^{\dagger}_{i+1} C \Psi_i,
\end{equation}
\begin{equation}
\begin{split}      
A=
\left(                 
  \begin{smallmatrix}   
    0,  & t_1 \\
    t_1, & 0
  \end{smallmatrix}
\right), B=
\left(                 
  \begin{smallmatrix}   
    J,  & 0 \\
    t_2, & J
  \end{smallmatrix}
\right),C=
\left(                 
  \begin{smallmatrix}   
    -J,  & t_2 \\
    0, & -J
  \end{smallmatrix}
\right).
\end{split}            
\end{equation}
Given the single particle state $|\Phi\rangle=\zeta_i \Psi^{\dagger}_i |0\rangle$, the Schrodinger equation can lead to the recursion function
\begin{equation}
A\zeta_i+B\zeta_{i-1}+C\zeta_{i+1}=\epsilon \zeta_i.
\end{equation}
Correspondingly, the transfer matrix  can be introduced  via
\begin{equation}
\begin{split}      
\left(                 
  \begin{smallmatrix}   
    \zeta_{i+1} \\ \zeta_{i}
  \end{smallmatrix}
\right)= T \left(                 
  \begin{smallmatrix}   
    \zeta_{i} \\ \zeta_{i-1}
  \end{smallmatrix}
\right)
\end{split}, T= \left(                 
  \begin{smallmatrix}   
    -C^{-1}(A-\epsilon I), & -C^{-1}B \\ I,  & 0
  \end{smallmatrix}
\right),        
\end{equation}
and the determinant of transfer matrix is
\begin{equation}
det(T)=det(C^{-1}(A-\epsilon I))*det((A-\epsilon I)^{-1}B)=1.
\end{equation}
Intuitively, the transfer matrix approach indicate that the  bulk states should  neither be boosted nor suppressed with the growing  lattice site index and  non-Hermitian skin effects can not be recognized.

At this stage, we tend to numerically demonstrate the density profiles of the representative bulk states. In Figs. 2(d-f), $|\psi^l_j|^2$ is plotted for  $l=1,10,20,50,51,100,91,81$, where $\psi^l_j$ denotes the $l$th eigenstate of the modified non-Hermitian SSH (NH-SSH) chain, $j$ being the site index. It is illustrated that the bulk  states  $|\psi^l_j|^2, $ are not equally extended, but are localized at the right and  left end of the lattice respectively, which suggests the presence of  non-Hermitian skin effects. Meanwhile, it shall be noticed that our system is in sharp contrast to the NH-SSH chain in ref [26], where for $t_1+\frac{r}{2}> (<) t_1-\frac{r}{2}$, all the bulk states are singly localized at the left (or right) end of the lattice. Besides, in the NH-SSH chain, the two zero energy end states collapse to the same one and become exceptional, exhibiting the single end boundary localization. Differently, specific to our model, it can be observed that $\psi^{50(51)}_j$ is the combination of the two zero energy edge states and has nearly equal localization on both end of the lattice. Hence, it can be identified that non-Hermitian skin effects are absent for the zero energy states.

Given the results above, we sum up that the modified NH-SSH chain possesses the non-Hermitian skin effect, featuring that the bulk states are piled up at different ends of the lattice. Meanwhile, such non-Hermitian skin effects are not continuous at the zero energy.  We tend to term this phenomenon as the discontinuous double-side non-Hermitian skin effects (DDS-NHSE), which is the critical physical origin for the phenomenon above. Remarkably, the DDS-NHSE can coexist with the preserved BBC.

Conventionally, non-Hermitian skin effects can also be decided by properties of the generalized Brillouin zone. In detail, the normal Bloch wave vector is replaced in the way that $e^{ik}\Leftrightarrow \beta $, through which the bulk iteration function of our modified NH-SSH chain can be depicted as
\begin{eqnarray}
\begin{aligned}
& J(\beta-\beta^{-1})\varphi_A+(t_2\beta^{-1}+t_1)\varphi_B=E \varphi_A, \\
& J(\beta-\beta^{-1})\varphi_B+(t_2\beta+t_1)\varphi_A=E \varphi_B,
\end{aligned}
\end{eqnarray}
where the cell wave function of the $n$th site is assumed to be  $\Psi_{n,A(B)}=\sum_j \beta^{n}_j \varphi^{j}_{A(B)}$. The topological edge modes are supposed to be boundary modes of zero energies, which will lead Eq. (11) to the following characteristic function
\begin{equation}
J^2 \beta^4-t_1t_2\beta^3-(2J^2+t^2_1+t^2_2)\beta^2-t_1t_2\beta +J^2=0.
\end{equation}
Eq. (12) is  quartic and has four roots $\beta_j, j=1,2,3,4$.  Meanwhile, the boundary cell wave functions satisfy
\begin{eqnarray}
\mathrm{left\,end}\left\{
\begin{aligned}
&t_1\Psi_{1,B}+J\Psi_{2,A}=E\Psi_{1,A}.\\
&t_1\Psi_{1,A}+t_2\Psi_{2,A}+J\Psi_{2,B}=E\Psi_{1,B}.
\end{aligned}
\right.
\end{eqnarray}
\begin{eqnarray}
\mathrm{right\,end}\left\{
\begin{aligned}
&-J\Psi_{N-1,B}+t_1\Psi_{N,A}=E\Psi_{N,B}. \\
&-J\Psi_{N-1,A}\!+\!t_2\Psi_{N-1,B}\!+\!t_1\Psi_{N,B}\!=\!E\Psi_{N,A}.
\end{aligned}
\right.
\end{eqnarray}
Through the equations above and according to the theory of generalized Brillouin zone, in the large size limit, two of the roots have to fulfill  $|\beta^{E\rightarrow0}_i|=|\beta^{E\rightarrow0}_j|, i\neq j$, which  will result in $|\beta^{E\rightarrow0}|=1$, suggesting the absence of  non-Hermitian skin effects at $E=0$ and the BBC should be preserved. These results exhibit a sound agreement with Fig. 2(a).  Meanwhile,  we have $|\beta^{E >(<) 0}_i|=|\beta^{E> (<)0}_j|\neq 1$ for $ \forall i \neq j$, which presents the non-Hermitian skin effects for states away from zero energy. Such discussions  are consistent with the analyses of DDS-NHSE.

\subsection{Restoring conventional BBC by introducing the extra non-Hermiticity}
In the previous discussion, it is shown that non-Hermitian skin effects can not guarantee the broken BBC. Now, we are to extend this result further and demonstrate the remarkable phenomenon that the modified BBC in Hermitian systems can be restored to be conventional by introducing external non-Hermiticity. For a detailed illustration, we still consider the modified NH-SSH chain and tune the purely imaginary inter-cell tunneling, $iJ$, to the general form $Je^{i\phi}$. In correspondence, the real and imaginary part of  energy spectrum take the following form respectively
\begin{eqnarray}
\begin{aligned}
&\mathrm{Re}[E(k)]=-2J\sin{\phi}\sin{(k)}\\
&\pm \sqrt{[t_1+t_2\cos{(k)}]^2+[t_2\sin{(k)}]^2}.
\end{aligned}
\end{eqnarray}
\begin{equation}
\mathrm{Im}[E(k)]=2J\cos{\phi}\sin{(k)}.
\end{equation}
For $\phi\neq N\pi$, $-2J\sin{\phi}\sin{(k)}$ in Eq.(15) can serve as $d_0(k) I$ and force the system to be semimetal. Consequently, based on the discussion in Hermitian cases, only  the direct band gap, which closes at $t_2=\pm 1.6 $ for $t_1=J=1,\phi=0.3\pi$, can properly describe the emergence of topological edge modes in open boundary conditions [see Fig. 3(a)].  Meanwhile, it can be observed that the minimum value of direct band gap amplitude $\mathrm{min}(|\delta|)$ is a function of $\phi$, and the variation of which can frequently manipulate the topological phase transition points [see Fig. 3(d)].

\begin{figure}[t]
\centering
\includegraphics[width=0.48\textwidth,height=0.29\textheight]{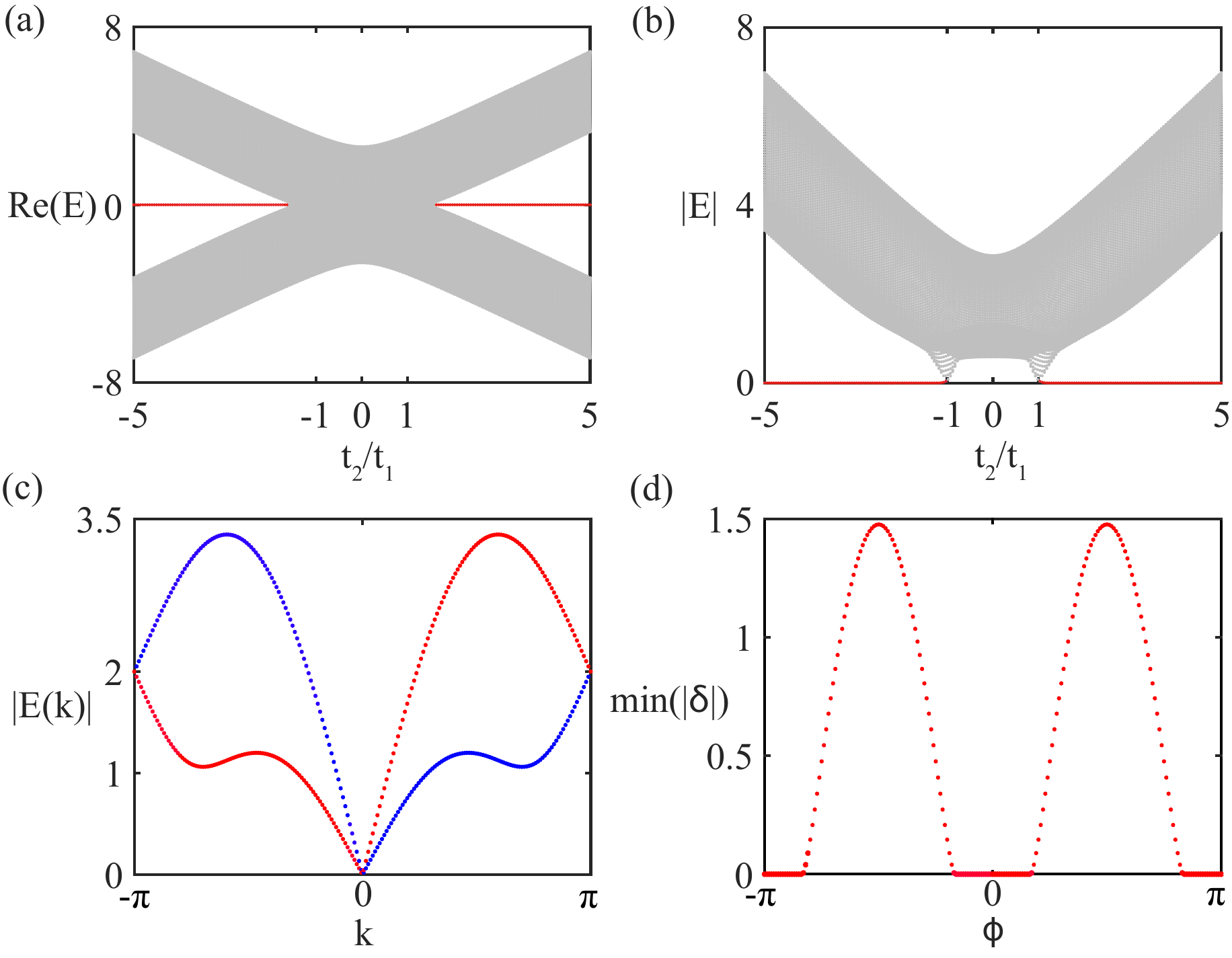}
\caption{(a) Considering the real part of energy spectrum with fixed $t_1=J=1.0,\phi=0.3\pi$, topological edge modes merge to the bulk at $t_2=\pm 1.6$, which is consistent with the discussion of modified BBC in Hermitian semimetals. (b) Considering the absolute energy spectrum in open boundary conditions, topological phase transitions take place with $|t_2|=|t_1|=1.0$. (c) The absolute energy spectrum in momentum space becomes gapless at $|t_2|=|t_1|$, suggesting  the preserved conventional BBC. (d) The minimum value of the direct band gap varies as a function of $\phi$. } \label{fig3}
\end{figure}

To proceed, we notice that the imaginary part of energy spectrum is gapless for all values of $k$, which can lead to the intuitive hypothesis that topological nontrivial regions shown by the absolute energy spectrum should coincide with those given by the real energy spectrum, both exhibiting modified BBC.

However, in Fig. 3(b), it is exhibited that the topological nontrivial edge modes presented by the absolute energy spectrum  merge to the bulk at $|t_2|=|t_1|$, where the momentum space bulk bands also just become gapless [see Fig. 3(c)]. Such nonintuitive results are equivalent to announce that non-Hermiticities are capable of restoring conventional BBC in Hermitian semimetals.

To illustrate the physical origin, it shall be kept in mind that different topological phases can also be distinguished by the information of line gap. For traditional Hermitian systems, energy spectrum can be flattened and projected to the real axis. In such cases,  line gap lies exactly on the imaginary axis.  Considering the topological trivial semimetal phases, different bands cross the line gap multiple times. Besides, as $t_2/t_1$ enlarges to the topological transition point, bands become separated and each band touches the line  gap  once [see Figs. 4(a-b)].  Nevertheless,  when the external non-Hermiticity (or imaginary eigen-energy) is  included,  a real line gap shall be changed to be complex and lie at other positions.  Specific to our cases, the localization of line gap should varies as a function of $\phi$ [see Figs. 4(c-d)]. Importantly, it should be noticed that as long as the line gap is not located at imaginary axis (i.e. any other positions), all two bands will touch the line gap only once with $|t_2|=|t_1|$ regardless of $\phi$, suggesting preserved BBC.  In other words, by including non-Hermiticity,  line gap can be tuned away from the imaginary axis, which is achieved by considering arbitrary $\phi,\phi\neq \frac{(2p+1)\pi}{2},p\in Z$, and any other general positions of the line gap are apt to restore the conventional BBC.

\subsection{Constraints on   $\mathbf{d_0(k)I}$ type preserved BBC}
Not all types of $d_0(k) I$ can result in nontrivial point gap topology, which fosters the non-Hermitian skin effects, and keep the topological  line gap  edge modes at the same time. In the following, we focus on the discussion of constraints.

\begin{figure}[t]
\centering
\includegraphics[width=0.48\textwidth,height=0.29\textheight]{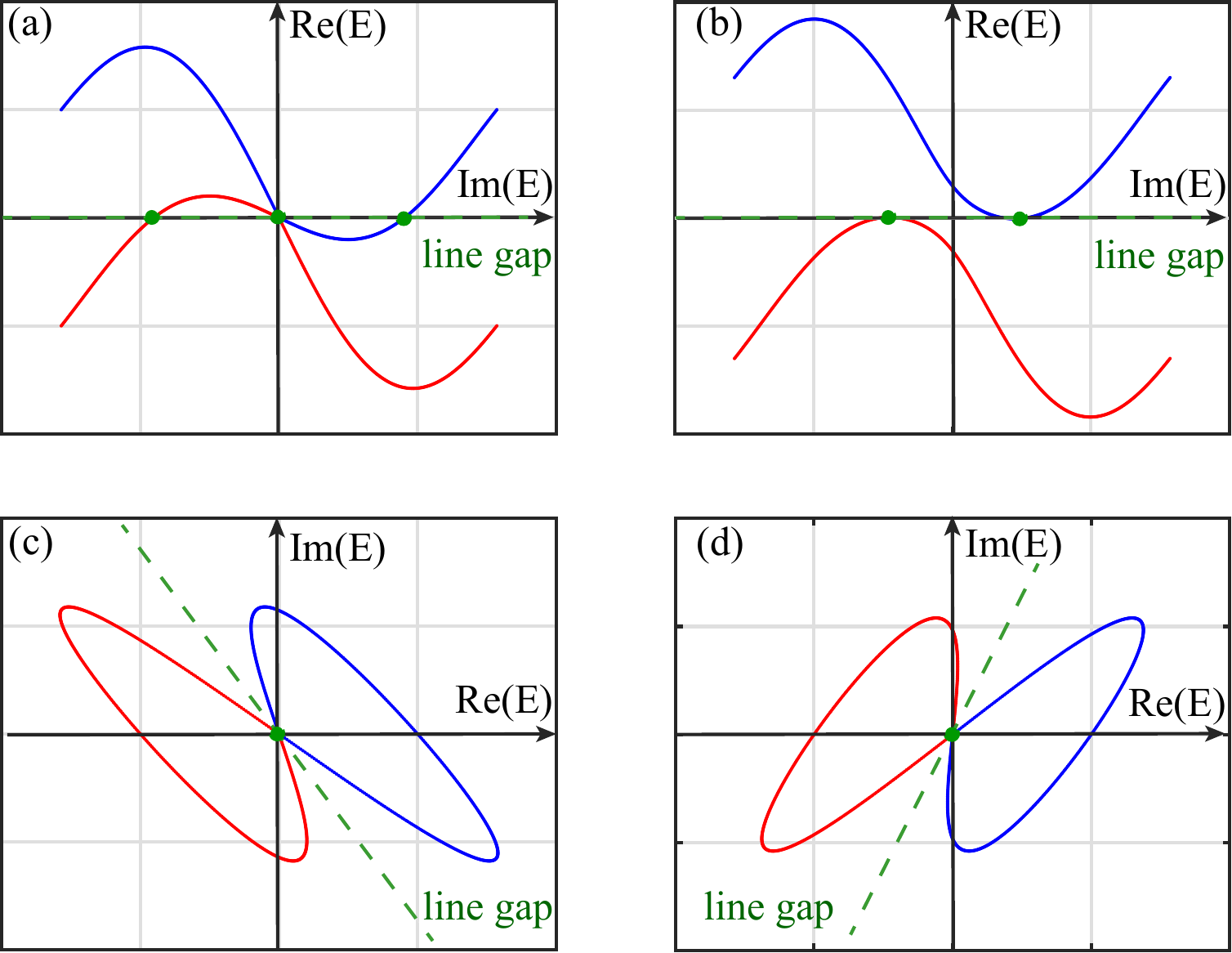}
\caption{(a-b) Only considering the real  energy spectrum, the line gap lies exactly on the imaginary axis. For $\phi=0.3\pi$, $t_1=1.0, J=1.0$, each band touches the line gap by multiple times (or once) for (a), $t_2=t_1=1.0$, semimetal phase. (or (b), $t_2=1.6\neq t_1$). Topological phase transitions take place in (b) with the upper band and lower band untouched. (c) Considering the complex energy spectrum, the line gaps do not lie on the imaginary axis but a general position. Preserved BBC is observed with $|t_2|=|t_1|=1.0,\phi=0.3\pi, J=1.0$. (d) The position of line gap changes as $\phi$ is tuned to be $0.8\pi$.} \label{fig4}
\end{figure}

First,  $d_0(k) I $ can hardly be compatible with the conventional chiral symmetry (CS$^\dagger$)\cite{PhysRevX.9.041015}.  Since $d_0(k) I$ will cause a energy shift to the upper and lower band in a same manner,  the chiral partner state of  $|u_n(k)\rangle$ with energy $-E_n(k)$ is not contained in the Hilbert space.  However, with the consideration of 38 fold symmetry ramification in non-Hermitian systems, purely imaginary $d_0(k) I$ can support CS symmetry, $\Gamma H^{\dagger} (k) \Gamma^{-1}=-H(k)$. In correspondence, we have band energies satisfying $E_n(k)=id_0(k)\pm \sqrt{d_x^2(k)+ d_y^2(k)+ d_z^2(k)}$ and real (imaginary) line gap exists for $d_x^2(k)+ d_y^2(k)+ d_z^2(k) >(<)0$. For the latter case, system can be purely anti-Hermitian and BBC is supposed to be reconstructed in a similar way as those shown in cases of semimetals.

For systems possessing TRS (TRS$^{\dagger}$) or PHS (PHS$^{\dagger}$ ) symmetry, which can be unified to the form $A H^{T}(k) A^{-1}= e^{i\Theta} H(-k)$ (or  $ A H^{*}(k) A^{-1}=e^{i\Theta} H(-k)$) and $A A^{\dagger}=1$, it is supposed to be fulfilled that $d_0(k)=e^{i\Theta} d_0(-k)$ (or $d^{*}_0(k)=e^{i\Theta} d_0(-k)$). Since $\Theta$ is irrelevant to $k$, the most common physical cases are $\Theta=2p\pi, (2p+1)\pi, p\in Z$ and the latter case shall force the spectrum to be central symmetric around $E=0$. Meanwhile, the condition $A A^{*}=-1$ can guarantee the  non-Hermitian Kramer's pairs and enable the possibilities of $Z_2$ skin effect. However, such conditions do not apply to the Hamiltonian in Eq. (5) and hence the DDS-NHSE differs from $Z_2$ skin effect.

In contrast to line gap topology, complex $d_0(k)I$ can greatly influence the point gap topology. As an illustration, considering a Hermitian Hamiltonian $H'$ free from skin effect, the synthesized Hamiltonian $H$, where $H=H'(k)+e^{i\gamma_p}d_0(k+\gamma_q)I$,  can exhibit the skin effect by properly choosing $\gamma_{p,q}$ to obtain a nontrivial phase difference. The Hamiltonian in Eq.(1) can serve as an example and another detailed demonstration is shown in appendix B. In consequence, the eigenenergy can be shifted in an artificial way by controlling the coefficients of $d_0(k)I$ and skin effects can be observed in a wide range of parameters. Therefore, it can be easily achieved  $|\beta^{E\rightarrow0}|=1$, which suggests the coexistence of skin effects and preserved bulk-boundary correspondence.

%
%

\section{Experimental Proposals}
To realize the modified NH-SSH chain, we propose to utilize the electrical circuits, which have been frequently applied in simulating various topological lattice models. Specifically, the experimental setups are shown  in  Fig. 5(a), where the blue and  red dots  are to distinguish $A, B$ types of sublattices respectively and the purple dashed line labels the unit cell.  The current flowing through the $j$th node of  the circuit lattice is  governed by the Kirchhoff's law
\begin{equation}
I_j=\sum_i L_{i,j} V_j.
\end{equation}
Here, $V_j$ depicts the voltage potential and $L_{i,j}$ is the net impedance of  all the electrical elements linked  to the node $j$, which is named as circuit Laplacian. Considering our systems,  the current on $A,B$ sublattices of the $j$th node is characterized by
\begin{eqnarray}
\begin{aligned}
I^{A}_j&=i\omega C_1[V^{A}_{j}-V^{B}_{j-1}]\!+\!i\omega C_2[V^A_{j}-V^B_{j}]\\
&+\frac{1}{i\omega L} V^{A}_{j}+ \frac{1}{R} (V^{A}_j-V^{A}_{j+1}),
\end{aligned}
\end{eqnarray}
\begin{eqnarray}
\begin{aligned}
I^B_{j}&=i\omega C_1[V^{B}_{j}-V^{A}_{j+1}]+i\omega C_2[V^B_{j}-V^A_{j}]\\
&+\frac{1}{i\omega L} V^{B}_{j}+\frac{1}{R} (V^{B}_j-V^{B}_{j+1}) ,
\end{aligned}
\end{eqnarray}
where $i\omega C_{1,2}$ are the impedances of the capacitors and $\frac{1}{i\omega L}$ is the impedance of the inductor.  $R$  measures the resistance of the operational amplifier, which serves as the negative impedance converter with current inversion (INIC), of which the details structure is shown in ref [50,52]. To be specific, for the current flowing towards the left side, $R=|R|$ is positive and for the rightward flowing current, $R=-|R|$ is negative.  The matrix form of the circuit Laplacian  can be written as
\begin{equation}L_{i,j}=
\begin{pmatrix}
  S & -i\omega C_2 & R^{-1} & 0 & 0 & ... \\
  -i \omega C_2  & S & -i\omega C_1& R^{-1} & 0 &  ... \\
  -R^{-1} & -i \omega C_1  & S & -i\omega C_2 & R^{-1} &...\\
  0 & -R^{-1} & -i \omega C_2 & S & -i \omega C_1\\
  ...
  \end{pmatrix},
\end{equation}
where the constant terms proportional to the identity matrix $I$ are of coefficient  $S=i\omega(C_1+C_2-\frac{1}{\omega^2 L})$. The impedances are related to parameters of the modified NH-SSH in the following way
\begin{equation}
t_1=-i \omega C_2, t_2=-i\omega C_1, J=R^{-1}.
\end{equation}
The resonant frequency of the electrical circuits can be obtained  as  $\bar{\omega}=1/\sqrt{L(C_1+ C_2)}$. The  presence of topological edge modes can be identified via the peaks of the two-points impedances between the left end node ($q_1$) and the right end mode ($q_2$),  which take the form $|Z^{\textrm {mod-SSH}}|=G(q_1,q_2)+G(q_2,q_1)-G(q_1,q_1)-G(q_2,q_2)$ and  $G$ is the circuit Green function. In Fig. 5(b), it is demonstrated that for a fixed large $|R|$, only when $ |C_1|> |C_2|$, the peaks of $|Z^{\textrm {mod-SSH}}|$ at $\omega=\bar{\omega}$ can be observed. Such results consist with the topological nontrivial region $|t_2|>|t_1|$.
\begin{figure}[t]
\centering
\includegraphics[width=0.49\textwidth,height=0.35\textheight]{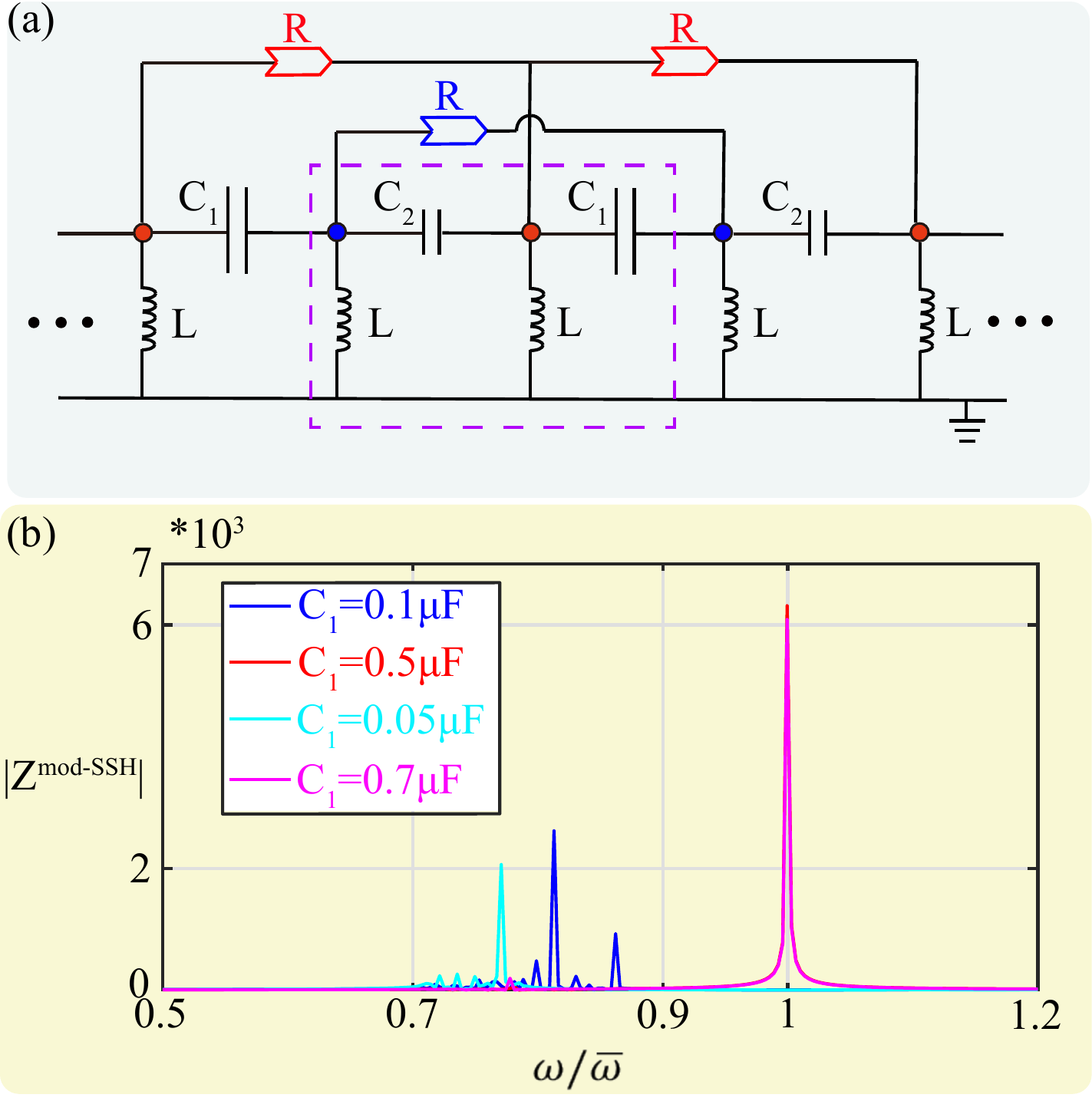}
\caption{(a) Realization of the modified NH-SSH chain via the electrical circuits. The purple dashed line labels a unit cell.  (b) The two-point impedance $|Z^{\textrm{mod-SSH}}|$ varies as a function of frequency $\omega/\bar{\omega}$ with fixed $ C_2=0.2\mu \mathrm{F},L=10 \mu \mathrm{H}, R=1.0*10^7 \Omega$. The peaks of $|Z^{\textrm{mod-SSH}}|$ are observed at $\omega=\bar{\omega}$ when $C_1=0.5\mu \mathrm{F}, 0.7\mu \mathrm{F}$, which coincide with the results of topological nontrivial regions.    } \label{fig5}
\end{figure}

\section{Conclusion}
We have demonstrated how the BBC  will be affected by the presence of momentum dependent energy shift, which takes the form of $d_0(k) I$.  It is observed that the  purely real $d_0(k) I$ in Hermitian systems can result in semimetal phases and the modified BBC should be reconstructed based on the definition of  direct band gaps.  In non-Hermitian cases, complex $d_0(k) I$ can greatly influence the point gap topology and facilitate the foundation of  skin effects. Remarkably, we propose the DDS-NHSE, which coexists with the preserved BBC. Meanwhile, contrary to the intuitive consideration, it is illustrated the conventional BBC  in Hermitian semimetals can be restored by introducing extra non-Hermiticity.  To the end, it shall be mentioned that $d_0(k) I$ can frequently appear in systems with the inter-cell tunneling between the same type of sublattices  or possessing the site-dependent chemical potential. Our work reveals the unexpected features of BBCs and has potential applications in constructing new types of non-Hermitian topological materials.

\section{Acknowledgement}
This work is supported by XRC-23079.

\appendix
\section{Analyzing bulk-boundary correspondence in the non-Hermitian SSH chain via transfer matrix approach.}
We tend to demonstrate how the transfer matrix approach  can succeed in predicting the skin effects via the example of  NH-SSH chain in ref [26]. In detail, the non-Hermiticities lie in the asymmetric  intra-cell tunneling, $t_1-\frac{r}{2}, t_1+\frac{r}{2}$ and  the inter-cell tunneling between the same type of sublattices is assumed to be absent. The Schrodinger equation of the bulk states corresponding to $A,B$ type of sublattices can be depicted by
\begin{eqnarray}
\begin{aligned}
&(t_1-\frac{r}{2})c_{j,A}+t_2c_{j+1,A}=E c_{j,B},\\
&(t_1+\frac{r}{2})c_{j,B}+t_2c_{j-1,B}=E c_{j,A},
\end{aligned}
\end{eqnarray}
where $c_{j,A}$ presents annihilating an $A$ type of particle at the $j$th site and $t_2$ is the inter-cell tunneling between different types of sublattices. $(t_1-(+)\ \frac{r}{2})$ depicts the rightwards (leftwards) tunneling.  Through Eq. (A1), the following iterative function can be obtained
\begin{eqnarray}
\begin{aligned}
&t_2[(t_1-\frac{r}{2})c_{j-1,B}+(t_1+\frac{r}{2})c_{j+1,B}]\\
&+(t^2_1+t^2_2-\frac{r^2}{4}-E^2)c_{j,B}=0,
\end{aligned}
\end{eqnarray}
and the transfer matrix is given as
\begin{equation}
\begin{split}
\left(                 
  \begin{matrix}   
      c_{j+1,B} \\
      c_{j,B}
  \end{matrix}
\right)=T \left(                 
  \begin{matrix}   
      c_{j,B} \\
      c_{j-1,B}
  \end{matrix}
\right),    
T\!=\!
\left(                 
  \begin{matrix}   
    a_{11},  & a_{12} \\
    1, & 0
  \end{matrix}
\right),
\end{split}
\end{equation}

\begin{equation}
a_{11}=-\frac{t^2_1+t^2_2-\frac{r^2}{4}-E^2}{t_2(t_1+\frac{r}{2})}, a_{12}=-\frac{t_1-\frac{r}{2}}{t_1+\frac{r}{2}}.
\end{equation}
The determinant of transfer matrix takes the form $|det(T)|=|a_{12}|=|\frac{t_1-\frac{r}{2}}{t_1+\frac{r}{2}} |$. Hence, when the tunneling towards the right hand side is larger than the tunneling towards the left hand side, $|t_1-\frac{r}{2}|>|t_1+\frac{r}{2}|$, we have $|det(T)|>1$ and all the states are piled up at the right end of the lattice. Respectively, $|det(T)|<1$ contributes to the left end localization. To sum up, whenever $|det(T)|\neq 1$, there shall be non-Hermitian skin effects and all states exhibit the boundary localization. For the cases that $|det(T)|=1$ and $|t_1-\frac{r}{2}|=|t_1+\frac{r}{2}|$, the system recovers Hermitian and non-Hermitian  skin effect vanishes.

\section{Controllable non-Hermitian skin effect in the monatomic quantum chain}

For the 1D monatomic chain, the most typical way to achieve skin effect comes via considering the Hatano-Nelson model, where the imbalanced leftward and rightward hopping ($|t_L|\neq |t_R|  $) can push the bulk states to one edge. Here, we propose another 1D structure with balanced hopping amplitude and the presence of skin effect can be artificially controlled. In detail, the Hamiltonian takes the form
\begin{equation}
H=\sum_j [te^{i\gamma_t}(c^{\dagger}_{j+1}c_j+c^{\dagger}_jc_{j+1})\!+\!\sum_p C(c^{\dagger}_{j+p}c_j+e^{i\gamma_c}c^{\dagger}_jc_{j+p}),
\end{equation}
where $C,t \in \mathrm{R}$ denote the quantum tunneling and $\gamma_{t(c)}$ are relevant phases, which can be manipulated by the techniques of laser assisted tunneling. The eigenenergy spectrum can be obtained as
\begin{equation}
E(k)=2te^{i\gamma_t}\cos{k}+ C[(e^{i\gamma_c}+1)\cos{kp}+i(e^{i\gamma_c}-1)\sin{kp}].
\end{equation}
First, we set $\gamma_t=m\pi,m\in Z$. The former term in Eq. (B2) is purely Hermitian and the latter term works in a similar way as the complex $d_0(k)I$.   For $\gamma_c=(2n+1)\pi, n\in Z$, the imaginary spectrum only contains $-2iC\sin{kp}$ and there are no terms proportional to $\cos{kp}$. In this case, $E(k)$ can form a closed loop in the complex plane regardless of value of $p$, which can give rise to  the skin effect. For $r_c=2n\pi, n\in Z$, all terms are real and the skin effects are absent.

However, there exists one special case, $\gamma_t\neq m\pi, \gamma_c=2n\pi$,  where  the real and imaginary part can coexist in the eigenenergy spectrum. Skin effects are not necessarily to be held.  To be specific, when $p=1$, $\mathrm{Re}E (k), \mathrm{Im} E(k)$ are of the equivalent phase and form a line with zero encircled area in the complex plane.  To form the non-equivalent phase and restore the skin effect, we can set $p\neq 1$. For this case,  skin effects are only decided by the length of effective tunneling, but not the asymmetric tunneling strength.
\bibliography{huanyu-2}

\begin{thebibliography}{53}%
\makeatletter
\providecommand \@ifxundefined [1]{%
 \@ifx{#1\undefined}
}%
\providecommand \@ifnum [1]{%
 \ifnum #1\expandafter \@firstoftwo
 \else \expandafter \@secondoftwo
 \fi
}%
\providecommand \@ifx [1]{%
 \ifx #1\expandafter \@firstoftwo
 \else \expandafter \@secondoftwo
 \fi
}%
\providecommand \natexlab [1]{#1}%
\providecommand \enquote  [1]{``#1''}%
\providecommand \bibnamefont  [1]{#1}%
\providecommand \bibfnamefont [1]{#1}%
\providecommand \citenamefont [1]{#1}%
\providecommand \href@noop [0]{\@secondoftwo}%
\providecommand \href [0]{\begingroup \@sanitize@url \@href}%
\providecommand \@href[1]{\@@startlink{#1}\@@href}%
\providecommand \@@href[1]{\endgroup#1\@@endlink}%
\providecommand \@sanitize@url [0]{\catcode `\\12\catcode `\$12\catcode
  `\&12\catcode `\#12\catcode `\^12\catcode `\_12\catcode `\%12\relax}%
\providecommand \@@startlink[1]{}%
\providecommand \@@endlink[0]{}%
\providecommand \url  [0]{\begingroup\@sanitize@url \@url }%
\providecommand \@url [1]{\endgroup\@href {#1}{\urlprefix }}%
\providecommand \urlprefix  [0]{URL }%
\providecommand \Eprint [0]{\href }%
\providecommand \doibase [0]{http://dx.doi.org/}%
\providecommand \selectlanguage [0]{\@gobble}%
\providecommand \bibinfo  [0]{\@secondoftwo}%
\providecommand \bibfield  [0]{\@secondoftwo}%
\providecommand \translation [1]{[#1]}%
\providecommand \BibitemOpen [0]{}%
\providecommand \bibitemStop [0]{}%
\providecommand \bibitemNoStop [0]{.\EOS\space}%
\providecommand \EOS [0]{\spacefactor3000\relax}%
\providecommand \BibitemShut  [1]{\csname bibitem#1\endcsname}%
\let\auto@bib@innerbib\@empty
\bibitem [{\citenamefont {Goldman}\ \emph {et~al.}(2010)\citenamefont
  {Goldman}, \citenamefont {Satija}, \citenamefont {Nikolic}, \citenamefont
  {Bermudez}, \citenamefont {Martin-Delgado}, \citenamefont {Lewenstein},\ and\
  \citenamefont {Spielman}}]{PhysRevLett.105.255302}%
  \BibitemOpen
  \bibfield  {author} {\bibinfo {author} {\bibfnamefont {N.}~\bibnamefont
  {Goldman}}, \bibinfo {author} {\bibfnamefont {I.}~\bibnamefont {Satija}},
  \bibinfo {author} {\bibfnamefont {P.}~\bibnamefont {Nikolic}}, \bibinfo
  {author} {\bibfnamefont {A.}~\bibnamefont {Bermudez}}, \bibinfo {author}
  {\bibfnamefont {M.~A.}\ \bibnamefont {Martin-Delgado}}, \bibinfo {author}
  {\bibfnamefont {M.}~\bibnamefont {Lewenstein}}, \ and\ \bibinfo {author}
  {\bibfnamefont {I.~B.}\ \bibnamefont {Spielman}},\ }\href@noop {} {\bibfield
  {journal} {\bibinfo  {journal} {Phys. Rev. Lett.}\ }\textbf {\bibinfo
  {volume} {105}},\ \bibinfo {pages} {255302} (\bibinfo {year}
  {2010})}\BibitemShut {NoStop}%
\bibitem [{\citenamefont {Zhang}\ \emph {et~al.}(2018)\citenamefont {Zhang},
  \citenamefont {Zhu}, \citenamefont {Zhao}, \citenamefont {Yan},\ and\
  \citenamefont {Zhu}}]{advances}%
  \BibitemOpen
  \bibfield  {author} {\bibinfo {author} {\bibfnamefont {D.-W.}\ \bibnamefont
  {Zhang}}, \bibinfo {author} {\bibfnamefont {Y.-Q.}\ \bibnamefont {Zhu}},
  \bibinfo {author} {\bibfnamefont {Y.}~\bibnamefont {Zhao}}, \bibinfo {author}
  {\bibfnamefont {H.}~\bibnamefont {Yan}}, \ and\ \bibinfo {author}
  {\bibfnamefont {S.-L.}\ \bibnamefont {Zhu}},\ }\href@noop {} {\bibfield
  {journal} {\bibinfo  {journal} {Adcances in Physics}\ }\textbf {\bibinfo
  {volume} {67}},\ \bibinfo {pages} {253} (\bibinfo {year} {2018})}\BibitemShut
  {NoStop}%
\bibitem [{\citenamefont {Goldman}\ \emph {et~al.}(2016)\citenamefont
  {Goldman}, \citenamefont {Budich},\ and\ \citenamefont {Zoller}}]{Natphys}%
  \BibitemOpen
  \bibfield  {author} {\bibinfo {author} {\bibfnamefont {N.}~\bibnamefont
  {Goldman}}, \bibinfo {author} {\bibfnamefont {J.}~\bibnamefont {Budich}}, \
  and\ \bibinfo {author} {\bibfnamefont {P.}~\bibnamefont {Zoller}},\
  }\href@noop {} {\bibfield  {journal} {\bibinfo  {journal} {Nat.Phys.}\
  }\textbf {\bibinfo {volume} {12}},\ \bibinfo {pages} {639} (\bibinfo {year}
  {2016})}\BibitemShut {NoStop}%
\bibitem [{\citenamefont {Liu}\ \emph {et~al.}(2013)\citenamefont {Liu},
  \citenamefont {Law}, \citenamefont {Ng},\ and\ \citenamefont
  {Lee}}]{PhysRevLett.111.120402}%
  \BibitemOpen
  \bibfield  {author} {\bibinfo {author} {\bibfnamefont {X.-J.}\ \bibnamefont
  {Liu}}, \bibinfo {author} {\bibfnamefont {K.~T.}\ \bibnamefont {Law}},
  \bibinfo {author} {\bibfnamefont {T.~K.}\ \bibnamefont {Ng}}, \ and\ \bibinfo
  {author} {\bibfnamefont {P.~A.}\ \bibnamefont {Lee}},\ }\href@noop {}
  {\bibfield  {journal} {\bibinfo  {journal} {Phys. Rev. Lett.}\ }\textbf
  {\bibinfo {volume} {111}},\ \bibinfo {pages} {120402} (\bibinfo {year}
  {2013})}\BibitemShut {NoStop}%
\bibitem [{\citenamefont {Laflamme}\ \emph {et~al.}(2014)\citenamefont
  {Laflamme}, \citenamefont {Baranov}, \citenamefont {Zoller},\ and\
  \citenamefont {Kraus}}]{PhysRevA.89.022319}%
  \BibitemOpen
  \bibfield  {author} {\bibinfo {author} {\bibfnamefont {C.}~\bibnamefont
  {Laflamme}}, \bibinfo {author} {\bibfnamefont {M.~A.}\ \bibnamefont
  {Baranov}}, \bibinfo {author} {\bibfnamefont {P.}~\bibnamefont {Zoller}}, \
  and\ \bibinfo {author} {\bibfnamefont {C.~V.}\ \bibnamefont {Kraus}},\
  }\href@noop {} {\bibfield  {journal} {\bibinfo  {journal} {Phys. Rev. A}\
  }\textbf {\bibinfo {volume} {89}},\ \bibinfo {pages} {022319} (\bibinfo
  {year} {2014})}\BibitemShut {NoStop}%
\bibitem [{\citenamefont {Sato}\ \emph {et~al.}(2009)\citenamefont {Sato},
  \citenamefont {Takahashi},\ and\ \citenamefont
  {Fujimoto}}]{PhysRevLett.103.020401}%
  \BibitemOpen
  \bibfield  {author} {\bibinfo {author} {\bibfnamefont {M.}~\bibnamefont
  {Sato}}, \bibinfo {author} {\bibfnamefont {Y.}~\bibnamefont {Takahashi}}, \
  and\ \bibinfo {author} {\bibfnamefont {S.}~\bibnamefont {Fujimoto}},\
  }\href@noop {} {\bibfield  {journal} {\bibinfo  {journal} {Phys. Rev. Lett.}\
  }\textbf {\bibinfo {volume} {103}},\ \bibinfo {pages} {020401} (\bibinfo
  {year} {2009})}\BibitemShut {NoStop}%
\bibitem [{\citenamefont {Chan}\ and\ \citenamefont
  {Gong}(2014)}]{PhysRevB.89.174501}%
  \BibitemOpen
  \bibfield  {author} {\bibinfo {author} {\bibfnamefont {C.~F.}\ \bibnamefont
  {Chan}}\ and\ \bibinfo {author} {\bibfnamefont {M.}~\bibnamefont {Gong}},\
  }\href@noop {} {\bibfield  {journal} {\bibinfo  {journal} {Phys. Rev. B}\
  }\textbf {\bibinfo {volume} {89}},\ \bibinfo {pages} {174501} (\bibinfo
  {year} {2014})}\BibitemShut {NoStop}%
\bibitem [{\citenamefont {Noh}\ \emph {et~al.}(2018)\citenamefont {Noh},
  \citenamefont {Benalcazar}, \citenamefont {Huang}, \citenamefont {Collins},
  \citenamefont {Chen}, \citenamefont {Hughes},\ and\ \citenamefont
  {Rechtsman}}]{Natpho}%
  \BibitemOpen
  \bibfield  {author} {\bibinfo {author} {\bibfnamefont {J.}~\bibnamefont
  {Noh}}, \bibinfo {author} {\bibfnamefont {W.~A.}\ \bibnamefont {Benalcazar}},
  \bibinfo {author} {\bibfnamefont {S.}~\bibnamefont {Huang}}, \bibinfo
  {author} {\bibfnamefont {M.~J.}\ \bibnamefont {Collins}}, \bibinfo {author}
  {\bibfnamefont {K.~P.}\ \bibnamefont {Chen}}, \bibinfo {author}
  {\bibfnamefont {T.~L.}\ \bibnamefont {Hughes}}, \ and\ \bibinfo {author}
  {\bibfnamefont {M.~C.}\ \bibnamefont {Rechtsman}},\ }\href@noop {} {\bibfield
   {journal} {\bibinfo  {journal} {Nat. Photonics}\ }\textbf {\bibinfo {volume}
  {12}},\ \bibinfo {pages} {408} (\bibinfo {year} {2018})}\BibitemShut
  {NoStop}%
\bibitem [{\citenamefont {Savelev}\ and\ \citenamefont
  {Gorlach}(2020)}]{PhysRevB.102.161112}%
  \BibitemOpen
  \bibfield  {author} {\bibinfo {author} {\bibfnamefont {R.~S.}\ \bibnamefont
  {Savelev}}\ and\ \bibinfo {author} {\bibfnamefont {M.~A.}\ \bibnamefont
  {Gorlach}},\ }\href@noop {} {\bibfield  {journal} {\bibinfo  {journal} {Phys.
  Rev. B}\ }\textbf {\bibinfo {volume} {102}},\ \bibinfo {pages} {161112}
  (\bibinfo {year} {2020})}\BibitemShut {NoStop}%
\bibitem [{\citenamefont {Szameit}\ \emph {et~al.}(2010)\citenamefont
  {Szameit}, \citenamefont {Dreisow}, \citenamefont {Heinrich}, \citenamefont
  {Keil}, \citenamefont {Nolte}, \citenamefont {T\"unnermann},\ and\
  \citenamefont {Longhi}}]{PhysRevLett.104.150403}%
  \BibitemOpen
  \bibfield  {author} {\bibinfo {author} {\bibfnamefont {A.}~\bibnamefont
  {Szameit}}, \bibinfo {author} {\bibfnamefont {F.}~\bibnamefont {Dreisow}},
  \bibinfo {author} {\bibfnamefont {M.}~\bibnamefont {Heinrich}}, \bibinfo
  {author} {\bibfnamefont {R.}~\bibnamefont {Keil}}, \bibinfo {author}
  {\bibfnamefont {S.}~\bibnamefont {Nolte}}, \bibinfo {author} {\bibfnamefont
  {A.}~\bibnamefont {T\"unnermann}}, \ and\ \bibinfo {author} {\bibfnamefont
  {S.}~\bibnamefont {Longhi}},\ }\href@noop {} {\bibfield  {journal} {\bibinfo
  {journal} {Phys. Rev. Lett.}\ }\textbf {\bibinfo {volume} {104}},\ \bibinfo
  {pages} {150403} (\bibinfo {year} {2010})}\BibitemShut {NoStop}%
\bibitem [{\citenamefont {Rechtsman}\ \emph
  {et~al.}(2013{\natexlab{a}})\citenamefont {Rechtsman}, \citenamefont
  {Zeuner}, \citenamefont {Plotnik}, \citenamefont {Lumer}, \citenamefont
  {Podolsky}, \citenamefont {Dreisow}, \citenamefont {Nolte}, \citenamefont
  {Segev},\ and\ \citenamefont {Szameit}}]{Nature}%
  \BibitemOpen
  \bibfield  {author} {\bibinfo {author} {\bibfnamefont {M.~C.}\ \bibnamefont
  {Rechtsman}}, \bibinfo {author} {\bibfnamefont {J.~M.}\ \bibnamefont
  {Zeuner}}, \bibinfo {author} {\bibfnamefont {Y.}~\bibnamefont {Plotnik}},
  \bibinfo {author} {\bibfnamefont {Y.}~\bibnamefont {Lumer}}, \bibinfo
  {author} {\bibfnamefont {D.}~\bibnamefont {Podolsky}}, \bibinfo {author}
  {\bibfnamefont {F.}~\bibnamefont {Dreisow}}, \bibinfo {author} {\bibfnamefont
  {S.}~\bibnamefont {Nolte}}, \bibinfo {author} {\bibfnamefont
  {M.}~\bibnamefont {Segev}}, \ and\ \bibinfo {author} {\bibfnamefont
  {A.}~\bibnamefont {Szameit}},\ }\href@noop {} {\bibfield  {journal} {\bibinfo
   {journal} {Nature}\ }\textbf {\bibinfo {volume} {496}},\ \bibinfo {pages}
  {196} (\bibinfo {year} {2013}{\natexlab{a}})}\BibitemShut {NoStop}%
\bibitem [{\citenamefont {Biswas}\ \emph {et~al.}(2021)\citenamefont {Biswas},
  \citenamefont {Dey},\ and\ \citenamefont {Ghosh}}]{PhysRevA.104.043513}%
  \BibitemOpen
  \bibfield  {author} {\bibinfo {author} {\bibfnamefont {P.}~\bibnamefont
  {Biswas}}, \bibinfo {author} {\bibfnamefont {S.}~\bibnamefont {Dey}}, \ and\
  \bibinfo {author} {\bibfnamefont {S.}~\bibnamefont {Ghosh}},\ }\href@noop {}
  {\bibfield  {journal} {\bibinfo  {journal} {Phys. Rev. A}\ }\textbf {\bibinfo
  {volume} {104}},\ \bibinfo {pages} {043513} (\bibinfo {year}
  {2021})}\BibitemShut {NoStop}%
\bibitem [{\citenamefont {Duggan}\ \emph {et~al.}(2020)\citenamefont {Duggan},
  \citenamefont {Mann},\ and\ \citenamefont {Al\`u}}]{PhysRevB.102.100303}%
  \BibitemOpen
  \bibfield  {author} {\bibinfo {author} {\bibfnamefont {R.}~\bibnamefont
  {Duggan}}, \bibinfo {author} {\bibfnamefont {S.~A.}\ \bibnamefont {Mann}}, \
  and\ \bibinfo {author} {\bibfnamefont {A.}~\bibnamefont {Al\`u}},\
  }\href@noop {} {\bibfield  {journal} {\bibinfo  {journal} {Phys. Rev. B}\
  }\textbf {\bibinfo {volume} {102}},\ \bibinfo {pages} {100303} (\bibinfo
  {year} {2020})}\BibitemShut {NoStop}%
\bibitem [{\citenamefont {Rechtsman}\ \emph
  {et~al.}(2013{\natexlab{b}})\citenamefont {Rechtsman}, \citenamefont
  {Plotnik}, \citenamefont {Zeuner}, \citenamefont {Song}, \citenamefont
  {Chen}, \citenamefont {Szameit},\ and\ \citenamefont
  {Segev}}]{PhysRevLett.111.103901}%
  \BibitemOpen
  \bibfield  {author} {\bibinfo {author} {\bibfnamefont {M.~C.}\ \bibnamefont
  {Rechtsman}}, \bibinfo {author} {\bibfnamefont {Y.}~\bibnamefont {Plotnik}},
  \bibinfo {author} {\bibfnamefont {J.~M.}\ \bibnamefont {Zeuner}}, \bibinfo
  {author} {\bibfnamefont {D.}~\bibnamefont {Song}}, \bibinfo {author}
  {\bibfnamefont {Z.}~\bibnamefont {Chen}}, \bibinfo {author} {\bibfnamefont
  {A.}~\bibnamefont {Szameit}}, \ and\ \bibinfo {author} {\bibfnamefont
  {M.}~\bibnamefont {Segev}},\ }\href@noop {} {\bibfield  {journal} {\bibinfo
  {journal} {Phys. Rev. Lett.}\ }\textbf {\bibinfo {volume} {111}},\ \bibinfo
  {pages} {103901} (\bibinfo {year} {2013}{\natexlab{b}})}\BibitemShut
  {NoStop}%
\bibitem [{\citenamefont {Lu}\ \emph {et~al.}(2021)\citenamefont {Lu},
  \citenamefont {Wang}, \citenamefont {Xiao}, \citenamefont {Zhang},\ and\
  \citenamefont {Chan}}]{PhysRevLett.126.113902}%
  \BibitemOpen
  \bibfield  {author} {\bibinfo {author} {\bibfnamefont {C.}~\bibnamefont
  {Lu}}, \bibinfo {author} {\bibfnamefont {C.}~\bibnamefont {Wang}}, \bibinfo
  {author} {\bibfnamefont {M.}~\bibnamefont {Xiao}}, \bibinfo {author}
  {\bibfnamefont {Z.~Q.}\ \bibnamefont {Zhang}}, \ and\ \bibinfo {author}
  {\bibfnamefont {C.~T.}\ \bibnamefont {Chan}},\ }\href@noop {} {\bibfield
  {journal} {\bibinfo  {journal} {Phys. Rev. Lett.}\ }\textbf {\bibinfo
  {volume} {126}},\ \bibinfo {pages} {113902} (\bibinfo {year}
  {2021})}\BibitemShut {NoStop}%
\bibitem [{\citenamefont {Benalcazar}\ \emph {et~al.}(2022)\citenamefont
  {Benalcazar}, \citenamefont {Noh}, \citenamefont {Wang}, \citenamefont
  {Huang}, \citenamefont {Chen},\ and\ \citenamefont
  {Rechtsman}}]{PhysRevB.105.195129}%
  \BibitemOpen
  \bibfield  {author} {\bibinfo {author} {\bibfnamefont {W.~A.}\ \bibnamefont
  {Benalcazar}}, \bibinfo {author} {\bibfnamefont {J.}~\bibnamefont {Noh}},
  \bibinfo {author} {\bibfnamefont {M.}~\bibnamefont {Wang}}, \bibinfo {author}
  {\bibfnamefont {S.}~\bibnamefont {Huang}}, \bibinfo {author} {\bibfnamefont
  {K.~P.}\ \bibnamefont {Chen}}, \ and\ \bibinfo {author} {\bibfnamefont
  {M.~C.}\ \bibnamefont {Rechtsman}},\ }\href@noop {} {\bibfield  {journal}
  {\bibinfo  {journal} {Phys. Rev. B}\ }\textbf {\bibinfo {volume} {105}},\
  \bibinfo {pages} {195129} (\bibinfo {year} {2022})}\BibitemShut {NoStop}%
\bibitem [{\citenamefont {Perczel}\ \emph {et~al.}(2020)\citenamefont
  {Perczel}, \citenamefont {Borregaard}, \citenamefont {Chang}, \citenamefont
  {Yelin},\ and\ \citenamefont {Lukin}}]{PhysRevLett.124.083603}%
  \BibitemOpen
  \bibfield  {author} {\bibinfo {author} {\bibfnamefont {J.}~\bibnamefont
  {Perczel}}, \bibinfo {author} {\bibfnamefont {J.}~\bibnamefont {Borregaard}},
  \bibinfo {author} {\bibfnamefont {D.~E.}\ \bibnamefont {Chang}}, \bibinfo
  {author} {\bibfnamefont {S.~F.}\ \bibnamefont {Yelin}}, \ and\ \bibinfo
  {author} {\bibfnamefont {M.~D.}\ \bibnamefont {Lukin}},\ }\href@noop {}
  {\bibfield  {journal} {\bibinfo  {journal} {Phys. Rev. Lett.}\ }\textbf
  {\bibinfo {volume} {124}},\ \bibinfo {pages} {083603} (\bibinfo {year}
  {2020})}\BibitemShut {NoStop}%
\bibitem [{\citenamefont {Daido}\ and\ \citenamefont
  {Yanase}(2019)}]{PhysRevB.100.174512}%
  \BibitemOpen
  \bibfield  {author} {\bibinfo {author} {\bibfnamefont {A.}~\bibnamefont
  {Daido}}\ and\ \bibinfo {author} {\bibfnamefont {Y.}~\bibnamefont {Yanase}},\
  }\href@noop {} {\bibfield  {journal} {\bibinfo  {journal} {Phys. Rev. B}\
  }\textbf {\bibinfo {volume} {100}},\ \bibinfo {pages} {174512} (\bibinfo
  {year} {2019})}\BibitemShut {NoStop}%
\bibitem [{\citenamefont {Tamura}\ \emph {et~al.}(2021)\citenamefont {Tamura},
  \citenamefont {Hoshino},\ and\ \citenamefont {Tanaka}}]{PhysRevB.104.165125}%
  \BibitemOpen
  \bibfield  {author} {\bibinfo {author} {\bibfnamefont {S.}~\bibnamefont
  {Tamura}}, \bibinfo {author} {\bibfnamefont {S.}~\bibnamefont {Hoshino}}, \
  and\ \bibinfo {author} {\bibfnamefont {Y.}~\bibnamefont {Tanaka}},\
  }\href@noop {} {\bibfield  {journal} {\bibinfo  {journal} {Phys. Rev. B}\
  }\textbf {\bibinfo {volume} {104}},\ \bibinfo {pages} {165125} (\bibinfo
  {year} {2021})}\BibitemShut {NoStop}%
\bibitem [{\citenamefont {Vu}(2022)}]{PhysRevB.105.064304}%
  \BibitemOpen
  \bibfield  {author} {\bibinfo {author} {\bibfnamefont {D.}~\bibnamefont
  {Vu}},\ }\href@noop {} {\bibfield  {journal} {\bibinfo  {journal} {Phys. Rev.
  B}\ }\textbf {\bibinfo {volume} {105}},\ \bibinfo {pages} {064304} (\bibinfo
  {year} {2022})}\BibitemShut {NoStop}%
\bibitem [{\citenamefont {Song}\ \emph {et~al.}(2020)\citenamefont {Song},
  \citenamefont {Elcoro},\ and\ \citenamefont {Bernevig}}]{Science}%
  \BibitemOpen
  \bibfield  {author} {\bibinfo {author} {\bibfnamefont {Z.-D.}\ \bibnamefont
  {Song}}, \bibinfo {author} {\bibfnamefont {L.}~\bibnamefont {Elcoro}}, \ and\
  \bibinfo {author} {\bibfnamefont {B.~A.}\ \bibnamefont {Bernevig}},\
  }\href@noop {} {\bibfield  {journal} {\bibinfo  {journal} {Science}\ }\textbf
  {\bibinfo {volume} {367}},\ \bibinfo {pages} {794} (\bibinfo {year}
  {2020})}\BibitemShut {NoStop}%
\bibitem [{\citenamefont {Song}\ \emph {et~al.}(2019)\citenamefont {Song},
  \citenamefont {Yao},\ and\ \citenamefont {Wang}}]{PhysRevLett.123.246801}%
  \BibitemOpen
  \bibfield  {author} {\bibinfo {author} {\bibfnamefont {F.}~\bibnamefont
  {Song}}, \bibinfo {author} {\bibfnamefont {S.}~\bibnamefont {Yao}}, \ and\
  \bibinfo {author} {\bibfnamefont {Z.}~\bibnamefont {Wang}},\ }\href@noop {}
  {\bibfield  {journal} {\bibinfo  {journal} {Phys. Rev. Lett.}\ }\textbf
  {\bibinfo {volume} {123}},\ \bibinfo {pages} {246801} (\bibinfo {year}
  {2019})}\BibitemShut {NoStop}%
\bibitem [{\citenamefont {Lin}\ \emph {et~al.}(2022)\citenamefont {Lin},
  \citenamefont {Li}, \citenamefont {Xiao}, \citenamefont {Wang}, \citenamefont
  {Yi},\ and\ \citenamefont {Xue}}]{PhysRevLett.129.113601}%
  \BibitemOpen
  \bibfield  {author} {\bibinfo {author} {\bibfnamefont {Q.}~\bibnamefont
  {Lin}}, \bibinfo {author} {\bibfnamefont {T.}~\bibnamefont {Li}}, \bibinfo
  {author} {\bibfnamefont {L.}~\bibnamefont {Xiao}}, \bibinfo {author}
  {\bibfnamefont {K.}~\bibnamefont {Wang}}, \bibinfo {author} {\bibfnamefont
  {W.}~\bibnamefont {Yi}}, \ and\ \bibinfo {author} {\bibfnamefont
  {P.}~\bibnamefont {Xue}},\ }\href@noop {} {\bibfield  {journal} {\bibinfo
  {journal} {Phys. Rev. Lett.}\ }\textbf {\bibinfo {volume} {129}},\ \bibinfo
  {pages} {113601} (\bibinfo {year} {2022})}\BibitemShut {NoStop}%
\bibitem [{\citenamefont {Xiao}\ \emph {et~al.}(2020)\citenamefont {Xiao},
  \citenamefont {Deng}, \citenamefont {Wang}, \citenamefont {Zhu},
  \citenamefont {Wang}, \citenamefont {Yi},\ and\ \citenamefont
  {Xue}}]{Natphys2}%
  \BibitemOpen
  \bibfield  {author} {\bibinfo {author} {\bibfnamefont {L.}~\bibnamefont
  {Xiao}}, \bibinfo {author} {\bibfnamefont {T.}~\bibnamefont {Deng}}, \bibinfo
  {author} {\bibfnamefont {K.}~\bibnamefont {Wang}}, \bibinfo {author}
  {\bibfnamefont {G.}~\bibnamefont {Zhu}}, \bibinfo {author} {\bibfnamefont
  {Z.}~\bibnamefont {Wang}}, \bibinfo {author} {\bibfnamefont {W.}~\bibnamefont
  {Yi}}, \ and\ \bibinfo {author} {\bibfnamefont {P.}~\bibnamefont {Xue}},\
  }\href@noop {} {\bibfield  {journal} {\bibinfo  {journal} {Nat. Phys.}\
  }\textbf {\bibinfo {volume} {16}},\ \bibinfo {pages} {761} (\bibinfo {year}
  {2020})}\BibitemShut {NoStop}%
\bibitem [{\citenamefont {Longhi}(2020)}]{PhysRevLett.124.066602}%
  \BibitemOpen
  \bibfield  {author} {\bibinfo {author} {\bibfnamefont {S.}~\bibnamefont
  {Longhi}},\ }\href@noop {} {\bibfield  {journal} {\bibinfo  {journal} {Phys.
  Rev. Lett.}\ }\textbf {\bibinfo {volume} {124}},\ \bibinfo {pages} {066602}
  (\bibinfo {year} {2020})}\BibitemShut {NoStop}%
\bibitem [{\citenamefont {Yao}\ and\ \citenamefont
  {Wang}(2018)}]{PhysRevLett.121.086803}%
  \BibitemOpen
  \bibfield  {author} {\bibinfo {author} {\bibfnamefont {S.}~\bibnamefont
  {Yao}}\ and\ \bibinfo {author} {\bibfnamefont {Z.}~\bibnamefont {Wang}},\
  }\href@noop {} {\bibfield  {journal} {\bibinfo  {journal} {Phys. Rev. Lett.}\
  }\textbf {\bibinfo {volume} {121}},\ \bibinfo {pages} {086803} (\bibinfo
  {year} {2018})}\BibitemShut {NoStop}%
\bibitem [{\citenamefont {Yi}\ and\ \citenamefont
  {Yang}(2020)}]{PhysRevLett.125.186802}%
  \BibitemOpen
  \bibfield  {author} {\bibinfo {author} {\bibfnamefont {Y.}~\bibnamefont
  {Yi}}\ and\ \bibinfo {author} {\bibfnamefont {Z.}~\bibnamefont {Yang}},\
  }\href@noop {} {\bibfield  {journal} {\bibinfo  {journal} {Phys. Rev. Lett.}\
  }\textbf {\bibinfo {volume} {125}},\ \bibinfo {pages} {186802} (\bibinfo
  {year} {2020})}\BibitemShut {NoStop}%
\bibitem [{\citenamefont {Zhang}\ \emph {et~al.}(2020)\citenamefont {Zhang},
  \citenamefont {Yang},\ and\ \citenamefont {Fang}}]{PhysRevLett.125.126402}%
  \BibitemOpen
  \bibfield  {author} {\bibinfo {author} {\bibfnamefont {K.}~\bibnamefont
  {Zhang}}, \bibinfo {author} {\bibfnamefont {Z.}~\bibnamefont {Yang}}, \ and\
  \bibinfo {author} {\bibfnamefont {C.}~\bibnamefont {Fang}},\ }\href@noop {}
  {\bibfield  {journal} {\bibinfo  {journal} {Phys. Rev. Lett.}\ }\textbf
  {\bibinfo {volume} {125}},\ \bibinfo {pages} {126402} (\bibinfo {year}
  {2020})}\BibitemShut {NoStop}%
\bibitem [{\citenamefont {Qi}\ \emph {et~al.}(2023)\citenamefont {Qi},
  \citenamefont {Han}, \citenamefont {Liu}, \citenamefont {Wang},\ and\
  \citenamefont {He}}]{PhysRevA.107.062214}%
  \BibitemOpen
  \bibfield  {author} {\bibinfo {author} {\bibfnamefont {L.}~\bibnamefont
  {Qi}}, \bibinfo {author} {\bibfnamefont {N.}~\bibnamefont {Han}}, \bibinfo
  {author} {\bibfnamefont {S.}~\bibnamefont {Liu}}, \bibinfo {author}
  {\bibfnamefont {H.-F.}\ \bibnamefont {Wang}}, \ and\ \bibinfo {author}
  {\bibfnamefont {A.-L.}\ \bibnamefont {He}},\ }\href@noop {} {\bibfield
  {journal} {\bibinfo  {journal} {Phys. Rev. A}\ }\textbf {\bibinfo {volume}
  {107}},\ \bibinfo {pages} {062214} (\bibinfo {year} {2023})}\BibitemShut
  {NoStop}%
\bibitem [{\citenamefont {Zhou}\ \emph {et~al.}(2023)\citenamefont {Zhou},
  \citenamefont {Jian}, \citenamefont {Pu}, \citenamefont {Lu}, \citenamefont
  {Huang}, \citenamefont {Deng}, \citenamefont {Ke},\ and\ \citenamefont
  {Liu}}]{Natcommu}%
  \BibitemOpen
  \bibfield  {author} {\bibinfo {author} {\bibfnamefont {Q.}~\bibnamefont
  {Zhou}}, \bibinfo {author} {\bibfnamefont {W.}~\bibnamefont {Jian}}, \bibinfo
  {author} {\bibfnamefont {Z.}~\bibnamefont {Pu}}, \bibinfo {author}
  {\bibfnamefont {J.}~\bibnamefont {Lu}}, \bibinfo {author} {\bibfnamefont
  {X.}~\bibnamefont {Huang}}, \bibinfo {author} {\bibfnamefont
  {W.}~\bibnamefont {Deng}}, \bibinfo {author} {\bibfnamefont {M.}~\bibnamefont
  {Ke}}, \ and\ \bibinfo {author} {\bibfnamefont {Z.}~\bibnamefont {Liu}},\
  }\href@noop {} {\bibfield  {journal} {\bibinfo  {journal} {Nat. Commun.}\
  }\textbf {\bibinfo {volume} {14}},\ \bibinfo {pages} {4569} (\bibinfo {year}
  {2023})}\BibitemShut {NoStop}%
\bibitem [{\citenamefont {Zhang}\ \emph {et~al.}(2021)\citenamefont {Zhang},
  \citenamefont {Tian}, \citenamefont {Jiang}, \citenamefont {Lu},\ and\
  \citenamefont {Chen}}]{Natcommu2}%
  \BibitemOpen
  \bibfield  {author} {\bibinfo {author} {\bibfnamefont {X.}~\bibnamefont
  {Zhang}}, \bibinfo {author} {\bibfnamefont {Y.}~\bibnamefont {Tian}},
  \bibinfo {author} {\bibfnamefont {J.-H.}\ \bibnamefont {Jiang}}, \bibinfo
  {author} {\bibfnamefont {M.-H.}\ \bibnamefont {Lu}}, \ and\ \bibinfo {author}
  {\bibfnamefont {Y.-F.}\ \bibnamefont {Chen}},\ }\href@noop {} {\bibfield
  {journal} {\bibinfo  {journal} {Nat. Commun.}\ }\textbf {\bibinfo {volume}
  {12}},\ \bibinfo {pages} {5377} (\bibinfo {year} {2021})}\BibitemShut
  {NoStop}%
\bibitem [{\citenamefont {Okuma}\ \emph {et~al.}(2020)\citenamefont {Okuma},
  \citenamefont {Kawabata}, \citenamefont {Shiozaki},\ and\ \citenamefont
  {Sato}}]{PhysRevLett.124.086801}%
  \BibitemOpen
  \bibfield  {author} {\bibinfo {author} {\bibfnamefont {N.}~\bibnamefont
  {Okuma}}, \bibinfo {author} {\bibfnamefont {K.}~\bibnamefont {Kawabata}},
  \bibinfo {author} {\bibfnamefont {K.}~\bibnamefont {Shiozaki}}, \ and\
  \bibinfo {author} {\bibfnamefont {M.}~\bibnamefont {Sato}},\ }\href@noop {}
  {\bibfield  {journal} {\bibinfo  {journal} {Phys. Rev. Lett.}\ }\textbf
  {\bibinfo {volume} {124}},\ \bibinfo {pages} {086801} (\bibinfo {year}
  {2020})}\BibitemShut {NoStop}%
\bibitem [{\citenamefont {Kawabata}\ \emph {et~al.}(2020)\citenamefont
  {Kawabata}, \citenamefont {Sato},\ and\ \citenamefont
  {Shiozaki}}]{PhysRevB.102.205118}%
  \BibitemOpen
  \bibfield  {author} {\bibinfo {author} {\bibfnamefont {K.}~\bibnamefont
  {Kawabata}}, \bibinfo {author} {\bibfnamefont {M.}~\bibnamefont {Sato}}, \
  and\ \bibinfo {author} {\bibfnamefont {K.}~\bibnamefont {Shiozaki}},\
  }\href@noop {} {\bibfield  {journal} {\bibinfo  {journal} {Phys. Rev. B}\
  }\textbf {\bibinfo {volume} {102}},\ \bibinfo {pages} {205118} (\bibinfo
  {year} {2020})}\BibitemShut {NoStop}%
\bibitem [{\citenamefont {Asjad}\ \emph {et~al.}(2023)\citenamefont {Asjad},
  \citenamefont {Teklu},\ and\ \citenamefont
  {Paris}}]{PhysRevResearch.5.013185}%
  \BibitemOpen
  \bibfield  {author} {\bibinfo {author} {\bibfnamefont {M.}~\bibnamefont
  {Asjad}}, \bibinfo {author} {\bibfnamefont {B.}~\bibnamefont {Teklu}}, \ and\
  \bibinfo {author} {\bibfnamefont {M.~G.~A.}\ \bibnamefont {Paris}},\
  }\href@noop {} {\bibfield  {journal} {\bibinfo  {journal} {Phys. Rev. Res.}\
  }\textbf {\bibinfo {volume} {5}},\ \bibinfo {pages} {013185} (\bibinfo {year}
  {2023})}\BibitemShut {NoStop}%
\bibitem [{\citenamefont {Mao}\ \emph {et~al.}(2023)\citenamefont {Mao},
  \citenamefont {Hao},\ and\ \citenamefont {Pan}}]{PhysRevA.107.043315}%
  \BibitemOpen
  \bibfield  {author} {\bibinfo {author} {\bibfnamefont {L.}~\bibnamefont
  {Mao}}, \bibinfo {author} {\bibfnamefont {Y.}~\bibnamefont {Hao}}, \ and\
  \bibinfo {author} {\bibfnamefont {L.}~\bibnamefont {Pan}},\ }\href@noop {}
  {\bibfield  {journal} {\bibinfo  {journal} {Phys. Rev. A}\ }\textbf {\bibinfo
  {volume} {107}},\ \bibinfo {pages} {043315} (\bibinfo {year}
  {2023})}\BibitemShut {NoStop}%
\bibitem [{\citenamefont {Guo}\ \emph {et~al.}(2021)\citenamefont {Guo},
  \citenamefont {Liu}, \citenamefont {Zhao}, \citenamefont {Liu},\ and\
  \citenamefont {Chen}}]{PhysRevLett.127.116801}%
  \BibitemOpen
  \bibfield  {author} {\bibinfo {author} {\bibfnamefont {C.-X.}\ \bibnamefont
  {Guo}}, \bibinfo {author} {\bibfnamefont {C.-H.}\ \bibnamefont {Liu}},
  \bibinfo {author} {\bibfnamefont {X.-M.}\ \bibnamefont {Zhao}}, \bibinfo
  {author} {\bibfnamefont {Y.}~\bibnamefont {Liu}}, \ and\ \bibinfo {author}
  {\bibfnamefont {S.}~\bibnamefont {Chen}},\ }\href@noop {} {\bibfield
  {journal} {\bibinfo  {journal} {Phys. Rev. Lett.}\ }\textbf {\bibinfo
  {volume} {127}},\ \bibinfo {pages} {116801} (\bibinfo {year}
  {2021})}\BibitemShut {NoStop}%
\bibitem [{\citenamefont {Richter}\ \emph {et~al.}(2017)\citenamefont
  {Richter}, \citenamefont {Michalsky}, \citenamefont {Sturm}, \citenamefont
  {Rosenow}, \citenamefont {Grundmann},\ and\ \citenamefont
  {Schmidt-Grund}}]{PhysRevA.95.023836}%
  \BibitemOpen
  \bibfield  {author} {\bibinfo {author} {\bibfnamefont {S.}~\bibnamefont
  {Richter}}, \bibinfo {author} {\bibfnamefont {T.}~\bibnamefont {Michalsky}},
  \bibinfo {author} {\bibfnamefont {C.}~\bibnamefont {Sturm}}, \bibinfo
  {author} {\bibfnamefont {B.}~\bibnamefont {Rosenow}}, \bibinfo {author}
  {\bibfnamefont {M.}~\bibnamefont {Grundmann}}, \ and\ \bibinfo {author}
  {\bibfnamefont {R.}~\bibnamefont {Schmidt-Grund}},\ }\href@noop {} {\bibfield
   {journal} {\bibinfo  {journal} {Phys. Rev. A}\ }\textbf {\bibinfo {volume}
  {95}},\ \bibinfo {pages} {023836} (\bibinfo {year} {2017})}\BibitemShut
  {NoStop}%
\bibitem [{\citenamefont {Bergholtz}\ \emph {et~al.}(2021)\citenamefont
  {Bergholtz}, \citenamefont {Budich},\ and\ \citenamefont
  {Kunst}}]{RevModPhys.93.015005}%
  \BibitemOpen
  \bibfield  {author} {\bibinfo {author} {\bibfnamefont {E.~J.}\ \bibnamefont
  {Bergholtz}}, \bibinfo {author} {\bibfnamefont {J.~C.}\ \bibnamefont
  {Budich}}, \ and\ \bibinfo {author} {\bibfnamefont {F.~K.}\ \bibnamefont
  {Kunst}},\ }\href@noop {} {\bibfield  {journal} {\bibinfo  {journal} {Rev.
  Mod. Phys.}\ }\textbf {\bibinfo {volume} {93}},\ \bibinfo {pages} {015005}
  (\bibinfo {year} {2021})}\BibitemShut {NoStop}%
\bibitem [{\citenamefont {Jin}\ \emph {et~al.}(2020)\citenamefont {Jin},
  \citenamefont {Wu}, \citenamefont {Wei},\ and\ \citenamefont
  {Song}}]{PhysRevB.101.045130}%
  \BibitemOpen
  \bibfield  {author} {\bibinfo {author} {\bibfnamefont {L.}~\bibnamefont
  {Jin}}, \bibinfo {author} {\bibfnamefont {H.~C.}\ \bibnamefont {Wu}},
  \bibinfo {author} {\bibfnamefont {B.-B.}\ \bibnamefont {Wei}}, \ and\
  \bibinfo {author} {\bibfnamefont {Z.}~\bibnamefont {Song}},\ }\href@noop {}
  {\bibfield  {journal} {\bibinfo  {journal} {Phys. Rev. B}\ }\textbf {\bibinfo
  {volume} {101}},\ \bibinfo {pages} {045130} (\bibinfo {year}
  {2020})}\BibitemShut {NoStop}%
\bibitem [{\citenamefont {Kawabata}\ \emph
  {et~al.}(2019{\natexlab{a}})\citenamefont {Kawabata}, \citenamefont
  {Bessho},\ and\ \citenamefont {Sato}}]{PhysRevLett.123.066405}%
  \BibitemOpen
  \bibfield  {author} {\bibinfo {author} {\bibfnamefont {K.}~\bibnamefont
  {Kawabata}}, \bibinfo {author} {\bibfnamefont {T.}~\bibnamefont {Bessho}}, \
  and\ \bibinfo {author} {\bibfnamefont {M.}~\bibnamefont {Sato}},\ }\href@noop
  {} {\bibfield  {journal} {\bibinfo  {journal} {Phys. Rev. Lett.}\ }\textbf
  {\bibinfo {volume} {123}},\ \bibinfo {pages} {066405} (\bibinfo {year}
  {2019}{\natexlab{a}})}\BibitemShut {NoStop}%
\bibitem [{\citenamefont {Yoshida}\ and\ \citenamefont
  {Hatsugai}(2023)}]{PhysRevB.107.075118}%
  \BibitemOpen
  \bibfield  {author} {\bibinfo {author} {\bibfnamefont {T.}~\bibnamefont
  {Yoshida}}\ and\ \bibinfo {author} {\bibfnamefont {Y.}~\bibnamefont
  {Hatsugai}},\ }\href@noop {} {\bibfield  {journal} {\bibinfo  {journal}
  {Phys. Rev. B}\ }\textbf {\bibinfo {volume} {107}},\ \bibinfo {pages}
  {075118} (\bibinfo {year} {2023})}\BibitemShut {NoStop}%
\bibitem [{\citenamefont {Xiao}\ \emph {et~al.}(2021)\citenamefont {Xiao},
  \citenamefont {Deng}, \citenamefont {Wang}, \citenamefont {Wang},
  \citenamefont {Yi},\ and\ \citenamefont {Xue}}]{PhysRevLett.126.230402}%
  \BibitemOpen
  \bibfield  {author} {\bibinfo {author} {\bibfnamefont {L.}~\bibnamefont
  {Xiao}}, \bibinfo {author} {\bibfnamefont {T.}~\bibnamefont {Deng}}, \bibinfo
  {author} {\bibfnamefont {K.}~\bibnamefont {Wang}}, \bibinfo {author}
  {\bibfnamefont {Z.}~\bibnamefont {Wang}}, \bibinfo {author} {\bibfnamefont
  {W.}~\bibnamefont {Yi}}, \ and\ \bibinfo {author} {\bibfnamefont
  {P.}~\bibnamefont {Xue}},\ }\href@noop {} {\bibfield  {journal} {\bibinfo
  {journal} {Phys. Rev. Lett.}\ }\textbf {\bibinfo {volume} {126}},\ \bibinfo
  {pages} {230402} (\bibinfo {year} {2021})}\BibitemShut {NoStop}%
\bibitem [{\citenamefont {Pap}\ \emph {et~al.}(2018)\citenamefont {Pap},
  \citenamefont {Boer},\ and\ \citenamefont {Waalkens}}]{PhysRevA.98.023818}%
  \BibitemOpen
  \bibfield  {author} {\bibinfo {author} {\bibfnamefont {E.~J.}\ \bibnamefont
  {Pap}}, \bibinfo {author} {\bibfnamefont {D.}~\bibnamefont {Boer}}, \ and\
  \bibinfo {author} {\bibfnamefont {H.}~\bibnamefont {Waalkens}},\ }\href@noop
  {} {\bibfield  {journal} {\bibinfo  {journal} {Phys. Rev. A}\ }\textbf
  {\bibinfo {volume} {98}},\ \bibinfo {pages} {023818} (\bibinfo {year}
  {2018})}\BibitemShut {NoStop}%
\bibitem [{\citenamefont {Solnyshkov}\ \emph {et~al.}(2021)\citenamefont
  {Solnyshkov}, \citenamefont {Leblanc}, \citenamefont {Bessonart},
  \citenamefont {Nalitov}, \citenamefont {Ren}, \citenamefont {Liao},
  \citenamefont {Li},\ and\ \citenamefont {Malpuech}}]{PhysRevB.103.125302}%
  \BibitemOpen
  \bibfield  {author} {\bibinfo {author} {\bibfnamefont {D.~D.}\ \bibnamefont
  {Solnyshkov}}, \bibinfo {author} {\bibfnamefont {C.}~\bibnamefont {Leblanc}},
  \bibinfo {author} {\bibfnamefont {L.}~\bibnamefont {Bessonart}}, \bibinfo
  {author} {\bibfnamefont {A.}~\bibnamefont {Nalitov}}, \bibinfo {author}
  {\bibfnamefont {J.}~\bibnamefont {Ren}}, \bibinfo {author} {\bibfnamefont
  {Q.}~\bibnamefont {Liao}}, \bibinfo {author} {\bibfnamefont {F.}~\bibnamefont
  {Li}}, \ and\ \bibinfo {author} {\bibfnamefont {G.}~\bibnamefont
  {Malpuech}},\ }\href@noop {} {\bibfield  {journal} {\bibinfo  {journal}
  {Phys. Rev. B}\ }\textbf {\bibinfo {volume} {103}},\ \bibinfo {pages}
  {125302} (\bibinfo {year} {2021})}\BibitemShut {NoStop}%
\bibitem [{\citenamefont {Yokomizo}\ and\ \citenamefont
  {Murakami}(2019)}]{PhysRevLett.123.066404}%
  \BibitemOpen
  \bibfield  {author} {\bibinfo {author} {\bibfnamefont {K.}~\bibnamefont
  {Yokomizo}}\ and\ \bibinfo {author} {\bibfnamefont {S.}~\bibnamefont
  {Murakami}},\ }\href@noop {} {\bibfield  {journal} {\bibinfo  {journal}
  {Phys. Rev. Lett.}\ }\textbf {\bibinfo {volume} {123}},\ \bibinfo {pages}
  {066404} (\bibinfo {year} {2019})}\BibitemShut {NoStop}%
\bibitem [{\citenamefont {Wang}\ \emph {et~al.}(2022)\citenamefont {Wang},
  \citenamefont {Zhao}, \citenamefont {Zhuang},\ and\ \citenamefont
  {Liu}}]{JPCM2}%
  \BibitemOpen
  \bibfield  {author} {\bibinfo {author} {\bibfnamefont {H.-Y.}\ \bibnamefont
  {Wang}}, \bibinfo {author} {\bibfnamefont {X.-M.}\ \bibnamefont {Zhao}},
  \bibinfo {author} {\bibfnamefont {L.}~\bibnamefont {Zhuang}}, \ and\ \bibinfo
  {author} {\bibfnamefont {W.-M.}\ \bibnamefont {Liu}},\ }\href@noop {}
  {\bibfield  {journal} {\bibinfo  {journal} {Journal of Physics: Condensed
  Matter}\ }\textbf {\bibinfo {volume} {34}},\ \bibinfo {pages} {365402}
  (\bibinfo {year} {2022})}\BibitemShut {NoStop}%
\bibitem [{\citenamefont {Longhi}(2019)}]{PhysRevResearch.1.023013}%
  \BibitemOpen
  \bibfield  {author} {\bibinfo {author} {\bibfnamefont {S.}~\bibnamefont
  {Longhi}},\ }\href@noop {} {\bibfield  {journal} {\bibinfo  {journal} {Phys.
  Rev. Res.}\ }\textbf {\bibinfo {volume} {1}},\ \bibinfo {pages} {023013}
  (\bibinfo {year} {2019})}\BibitemShut {NoStop}%
\bibitem [{\citenamefont {Zou}\ \emph {et~al.}(2021)\citenamefont {Zou},
  \citenamefont {Chen}, \citenamefont {He}, \citenamefont {Bao}, \citenamefont
  {Lee}, \citenamefont {Sun},\ and\ \citenamefont {Zhang}}]{Natcommu3}%
  \BibitemOpen
  \bibfield  {author} {\bibinfo {author} {\bibfnamefont {D.}~\bibnamefont
  {Zou}}, \bibinfo {author} {\bibfnamefont {T.}~\bibnamefont {Chen}}, \bibinfo
  {author} {\bibfnamefont {W.}~\bibnamefont {He}}, \bibinfo {author}
  {\bibfnamefont {J.}~\bibnamefont {Bao}}, \bibinfo {author} {\bibfnamefont
  {C.-H.}\ \bibnamefont {Lee}}, \bibinfo {author} {\bibfnamefont
  {H.}~\bibnamefont {Sun}}, \ and\ \bibinfo {author} {\bibfnamefont
  {X.}~\bibnamefont {Zhang}},\ }\href@noop {} {\bibfield  {journal} {\bibinfo
  {journal} {Nat. Commun.}\ }\textbf {\bibinfo {volume} {12}},\ \bibinfo
  {pages} {7201} (\bibinfo {year} {2021})}\BibitemShut {NoStop}%
\bibitem [{\citenamefont {Wang}\ and\ \citenamefont
  {Liu}(2022)}]{PhysRevA.106.052216}%
  \BibitemOpen
  \bibfield  {author} {\bibinfo {author} {\bibfnamefont {H.-Y.}\ \bibnamefont
  {Wang}}\ and\ \bibinfo {author} {\bibfnamefont {W.-M.}\ \bibnamefont {Liu}},\
  }\href@noop {} {\bibfield  {journal} {\bibinfo  {journal} {Phys. Rev. A}\
  }\textbf {\bibinfo {volume} {106}},\ \bibinfo {pages} {052216} (\bibinfo
  {year} {2022})}\BibitemShut {NoStop}%
\bibitem [{\citenamefont {Ezawa}(2019)}]{PhysRevB.100.045407}%
  \BibitemOpen
  \bibfield  {author} {\bibinfo {author} {\bibfnamefont {M.}~\bibnamefont
  {Ezawa}},\ }\href@noop {} {\bibfield  {journal} {\bibinfo  {journal} {Phys.
  Rev. B}\ }\textbf {\bibinfo {volume} {100}},\ \bibinfo {pages} {045407}
  (\bibinfo {year} {2019})}\BibitemShut {NoStop}%
\bibitem [{\citenamefont {Imhof}\ \emph {et~al.}(2018)\citenamefont {Imhof},
  \citenamefont {Berger}, \citenamefont {Bayer}, \citenamefont {Brehm},
  \citenamefont {Molenkamp}, \citenamefont {Kiessling}, \citenamefont
  {Schindler}, \citenamefont {Lee}, \citenamefont {Greiter}, \citenamefont
  {Neupert},\ and\ \citenamefont {Thomale}}]{Natphys3}%
  \BibitemOpen
  \bibfield  {author} {\bibinfo {author} {\bibfnamefont {S.}~\bibnamefont
  {Imhof}}, \bibinfo {author} {\bibfnamefont {C.}~\bibnamefont {Berger}},
  \bibinfo {author} {\bibfnamefont {F.}~\bibnamefont {Bayer}}, \bibinfo
  {author} {\bibfnamefont {J.}~\bibnamefont {Brehm}}, \bibinfo {author}
  {\bibfnamefont {L.}~\bibnamefont {Molenkamp}}, \bibinfo {author}
  {\bibfnamefont {T.}~\bibnamefont {Kiessling}}, \bibinfo {author}
  {\bibfnamefont {F.}~\bibnamefont {Schindler}}, \bibinfo {author}
  {\bibfnamefont {C.}~\bibnamefont {Lee}}, \bibinfo {author} {\bibfnamefont
  {M.}~\bibnamefont {Greiter}}, \bibinfo {author} {\bibfnamefont
  {T.}~\bibnamefont {Neupert}}, \ and\ \bibinfo {author} {\bibfnamefont
  {R.}~\bibnamefont {Thomale}},\ }\href@noop {} {\bibfield  {journal} {\bibinfo
   {journal} {Nat.Phys.}\ }\textbf {\bibinfo {volume} {14}},\ \bibinfo {pages}
  {925} (\bibinfo {year} {2018})}\BibitemShut {NoStop}%
\bibitem [{\citenamefont {Hofmann}\ \emph {et~al.}(2019)\citenamefont
  {Hofmann}, \citenamefont {Helbig}, \citenamefont {Lee}, \citenamefont
  {Greiter},\ and\ \citenamefont {Thomale}}]{PhysRevLett.122.247702}%
  \BibitemOpen
  \bibfield  {author} {\bibinfo {author} {\bibfnamefont {T.}~\bibnamefont
  {Hofmann}}, \bibinfo {author} {\bibfnamefont {T.}~\bibnamefont {Helbig}},
  \bibinfo {author} {\bibfnamefont {C.~H.}\ \bibnamefont {Lee}}, \bibinfo
  {author} {\bibfnamefont {M.}~\bibnamefont {Greiter}}, \ and\ \bibinfo
  {author} {\bibfnamefont {R.}~\bibnamefont {Thomale}},\ }\href@noop {}
  {\bibfield  {journal} {\bibinfo  {journal} {Phys. Rev. Lett.}\ }\textbf
  {\bibinfo {volume} {122}},\ \bibinfo {pages} {247702} (\bibinfo {year}
  {2019})}\BibitemShut {NoStop}%
\bibitem [{\citenamefont {Kawabata}\ \emph
  {et~al.}(2019{\natexlab{b}})\citenamefont {Kawabata}, \citenamefont
  {Shiozaki}, \citenamefont {Ueda},\ and\ \citenamefont
  {Sato}}]{PhysRevX.9.041015}%
  \BibitemOpen
  \bibfield  {author} {\bibinfo {author} {\bibfnamefont {K.}~\bibnamefont
  {Kawabata}}, \bibinfo {author} {\bibfnamefont {K.}~\bibnamefont {Shiozaki}},
  \bibinfo {author} {\bibfnamefont {M.}~\bibnamefont {Ueda}}, \ and\ \bibinfo
  {author} {\bibfnamefont {M.}~\bibnamefont {Sato}},\ }\href@noop {} {\bibfield
   {journal} {\bibinfo  {journal} {Phys. Rev. X}\ }\textbf {\bibinfo {volume}
  {9}},\ \bibinfo {pages} {041015} (\bibinfo {year}
  {2019}{\natexlab{b}})}\BibitemShut {NoStop}%
\end{thebibliography}%
\end{document}